\documentclass[aps,prb,twocolumn]{revtex4}
\usepackage{bm}
\usepackage{graphicx}
\usepackage{amsmath}

\newcommand{\BEQ}{\begin{equation}}
\newcommand{\EEQ}{\end{equation}}
\newcommand{\BEA}{\begin{eqnarray}}
\newcommand{\EEA}{\end{eqnarray}}
 
\newcommand{\p}{\partial}
\newcommand{\nn}{\nonumber }
\newcommand{\rpos}{{\bf r}}
\newcommand{\bfr}{{\bf r}}
\makeatletter
\newcommand\figcaption{\def\@captype{figure}\caption}
\makeatother

\begin{document}

\title{Complexity of waves in nonlinear disordered media}

\author{C. Conti$^{1,3}$, L. Leuzzi$^{2,3}$} 
\email{luca.leuzzi@cnr.it, claudio.conti@roma1.infn.it}
\affiliation{ $^1$ ISC-CNR, UOS Roma, Piazzale A. Moro 2,
 I-00185, Roma, Italy \\ $^2$ IPCF-CNR, UOS Roma, Piazzale A. Moro 2,
 I-00185, Roma, Italy \\ $^3$ Dipartimento di Fisica, Universit\`a di
 Roma ``Sapienza,'' Piazzale A. Moro 2, I-00185, Roma, Italy}

\begin{abstract}
The statistical properties of the phases of several modes nonlinearly
coupled in a random system are investigated by means of a Hamiltonian
model with disordered couplings.  The regime in which the modes have a
stationary distribution of their energies and the phases are coupled
is studied for arbitrary degrees of randomness and energy. The
complexity versus temperature and strength of nonlinearity is
calculated.  A phase diagram is derived in terms of the stored energy
and amount of disorder.  Implications in random lasing, nonlinear wave
propagation and finite temperature Bose-Einstein condensation are
discussed.
\end{abstract}

\maketitle

The interplay between disorder and nonlinearity in wave-propagation is
a technically challenging process.  Such a problem arises in several
different frameworks in modern physics, as nonlinear optical
propagation and laser emission in random systems, Bose-Einstein
condensation (BEC) and Anderson localization (as, e.g., in
Refs. [\onlinecite{Swartz07, InguscioNature2008,
AspectNature2008,Conti07, Kivshar10, Wiersma08,Ciuti10,Bertolozzo10,
Kaiser10, Shapiro10, 
Conti05Complex, Conti06Complex, Folli10,Skipetrov03, Zaitsev09,Cao03r,Tureci:08,Kaiser10,vandermolen:06b,Lag2010}]). Related
topics are the super-continuum generation and condensation
processes. \cite{Conti08, Weill10, Gat10, Picozzi10, Bortolozzo09,
Turitsyn10}
\\
\indent
When disorder has a leading role, nonlinear processes can be largely
hampered as due to the fact that waves rapidly diffuse in the system.
Conversely, if the structural disorder is perturbative, its effect on
nonlinear evolution is typically marginal, leading to some
additional linear or nonlinear scattering-losses, but not radically
affecting the qualitative nonlinear regime expected in the absence of
disorder.
When disorder and nonlinearity play on the same ground, one can
envisage novel and fascinating physical phenomena; however, the technical
analysis is rather difficult, as the problem cannot be attacked by
perturbational expansions.
\\
\indent
Physically, disorder and nonlinearity compete in those regimes when
wave scattering affects the degree of localization, eventually
inducing it (as in the Anderson localization), and nonlinearity
couples the modes in the system. These may in general exhibit a
distribution of localization lengths (determined by the amount of
disorder) and a strength of the interaction depending on the amount of
energy coupled in the system. 
\\ \indent Our interest here is to provide a general theoretical
framework, whose result is the prediction of specific transitions from
incoherent to coherent regimes, which are specifically due to the
disorder and display a glassy character, associated with a large
number of degenerate states present in the system.
\\ \indent We adopt a statistical mechanics perspective to the
problem, which allows to derive very general conclusions, not
depending on the specific problem, and our focus is on the case in
which many modes are excited.  This implies that energy is distributed
among many excitations in an initial stage of the dynamics.  The
overall coherence (i.e., the statistical properties of the overall
wave) will be determined by the phase-relations between the involved
modes. Here we show that there exist collective disordered regimes,
where coherence is dictated by the fact that the system is trapped in
one of many energetically equivalent states, as described below.
\\ \indent Representing mode phases by means of continuous planar
XY-like spins and applying a statistical mechanic approach we can
identify different thermodynamic phases. For negligible nonlinearity,
all the modes will oscillate independently in a continuous wave noisy
regime (``paramagnetic''-like phase).  For a strong interaction and a
suitable sign of the nonlinear coefficients, all the modes will
oscillate coherently (``ferromagnetic''-like regime). This
corresponds, for example, to standard passively mode-locked laser
systems \cite{Haus00}, which we found to take place even in the
presence of a certain amount of disorder.  In intermediate regimes,
the tendency to oscillate synchronously will be frustrated by
disorder, resulting in a glassy regime.
\\ \indent These three regimes are identified by a set of order
parameters (up to $10$ for the most complicated phase, as detailed
below), which can be cast into two classes: the ``magnetizations''
$m$, and the ``overlaps'' $q$. As the system is in the
paramagnetic-like phase all $m$ and $q$ vanish; in the ferromagnetic
regime they are both different from zero; while in the glassy phase
$m=0$ and (at least some of) the overlap parameters are different from
zero.
\\ \indent The paramagnetic and ferromagnetic phases may be present
even in the absence of disorder; conversely, a necessary condition to
find a glassy phase is frustration (disorder induced in our case, see
Sec. \ref{sec:frustration}). The glassy phase is characterized by the
occurrence of a rugged - {\rm complex} - landscape for the Gibbs free
energy functional in the mode phases space: a huge number of minima
are present, corresponding to a multitude of stable and metastable
states in the system, separated by barriers of various heights and
clustered in basins.  This is a result of the competition between
disorder and nonlinearity.
\\ \indent The existence of a not-vanishing {\em complexity} (which
measures the number of energetically equivalent states) for the
possible distributions of mode-phases is the basic ingredient for
explaining a variety of novel phenomena like speckle pattern
fluctuations and spectral statistics for disordered, or weakly
disordered nonlinear, systems, ergodicity breaking, glassy transitions
of light or BEC, and ultimately the onset of a coherent regime in a
random nonlinear system.
\\ \indent Our work extends previously reported results,
cf. Ref. [\onlinecite{Leuzzi09}] and it includes an arbitrary degree
of disorder and the discussion of its application to nonlinear
Schroedinger models, relevant, e.g., for BEC, spatial nonlinear optics
and supercontinuum generation.
\\
\indent
The paper is organized as follows: in Sec. \ref{sec:model} we
introduce the model and we discuss some of its possible fields of application, namely random lasers, Bose-Einstein condensates, optical
propagation; in Sec. \ref{sec:random} we discuss
the effect of disorder in the coupling of light modes and the new
expected phenomena; we dedicate Sec. \ref{sec:statmech} to an extremely
basic introduction to the statistical mechanics of systems with
quenched disorder, to the replica method, and to the definition of 
complexity; in Sec. \ref{sec:computation} we study the
model within the replica approach, details of the computation are
reported in App. \ref{app}; in Sec. \ref{sec:complexity} we
discuss the presence of excited metastable states and we compute the
complexity functional; in Sec. \ref{sec:phdi} we show the phase
diagrams of our model and discuss the properties of its thermodynamic
phases; eventually, in Sec. \ref{sec:conclusion} we draw our
conclusions.

\section{The leading model}
\label{sec:model}
Here we review some of the disordered systems where a relevant non
linear interaction may arise and our model applies.  The basic
Hamiltonian of $N$ adimensional angular variables $\phi\in [0:2\pi]$
is given by
\begin{equation}
{\cal H}_J[\phi]=
-\hspace*{-.5cm}\sum_{\footnotesize{\begin{array}{c} i_1<\!i_2, i_3<\!i_4
\\ 
i_1<\!i_3 \end{array}}}^{1,N}\hspace*{-.5cm}J_{\bf{i}}
\cos(\phi_{i_1}+\phi_{i_2}-\phi_{i_3}-\phi_{i_4})
\label{mainHam}
\end{equation}
where ${\bf i}=\{i_1,i_2,i_3i_4\}$ and $J_{\bf i}$ are random
independent identically distributed interaction variables. Formally,
the couplings can vary from short- to long-range, depending on the
structure of the four-index interaction tensor $J_{\bf i}$.  If we
choose $J_{\bf i}\neq 0$ for any distinct quadruple $i_1, \ldots,
i_4$, independently of the geometric position, we can build a
mean-field theory in which the system is fully connected. In this case
the average $J_{\bm i}$ and the variance of its distribution must
scale as $1/N^3$ to guarantee thermodynamic convergence of (free)
energy density.  The interaction can, otherwise, be bond-diluted with
arbitrary degree of diluteness, adopting a sparse tensor whose
non-zero elements do not scale with the number of modes,
\\
\indent
As we show in the following, the Hamiltonian, Eq. (\ref{mainHam}) is
derived in different contexts, and the various parameters may have
different interpretation.  In this manuscript we want to derive general
properties that are expected assuming a simple, yet reasonable,
Gaussian distribution for the random coupling coefficients, with a
non-vanishing mean value. Varying the ratio between the standard
deviation and the mean value we control a different degree of
disorder. Hence, these results applies to the various cases in which
random wave propagation, localization and not-negligible nonlinear
effects are important; a few of them are detailed in the following.
\\ \indent As a thermodynamic approach is adopted, one can argue if
the statistical mechanic techniques also apply in those systems where
the definition of a temperature is not straightforward, as,
specifically, nonlinear optical wave propagation in disordered media.
This particular problem can, then, be treated as for constraint
satisfaction problems in computer science,
\cite{Kirkpatrick94,Monasson99,Mezard02,Mezard09} where - at the end
of the calculation - the limit of zero-temperature is taken and it is
shown that a transition is expected as the number of constraints grows.

\subsection{Random active and passive electromagnetic cavities}
\label{ss:rl}
We start from the electromagnetic energy inside a dielectric cavity
(due to the generality of the considered model similar examples can be
found in a variety of systems):
\begin{equation}
\mathcal{E}_{EM}=\int {\bf E}(\bm r)\cdot {\bf D}(\bm r)~ dV
\end{equation}
The displacement vector is written in terms of a position dependent
relative dielectric constant $\epsilon_r({\bf r})$:
\begin{equation}
{\bf D}({\bf r})=\epsilon_0 \epsilon_r({\bf r}) {\bf E}({\bf
r})+\epsilon_0 {\bf P}_{NL}({\bf r})
\end{equation}
with ${\bf P}_{NL}$ the nonlinear polarization. In absence of the
latter, for a closed cavity, the field can be expanded in terms of the
modes of the system. In the presence of disorder these modes may
display a different degree of localization as, e.g.,  in a
disordered photonic crystal (PhC). \cite{Conti08PhC} For a closed cavity
these modes form a complete set and the field can be expanded in terms
of the modes
\begin{equation}
{\bf E}=\Re \left[\sum_{n=1}^N 
a_n(t) \exp{\left(-\imath\omega_n t\right)} {\bf E}_n(\rpos)\right]
\label{modeexpansion}
\end{equation}
with ${\bf E}=\{E^x,E^y,E^z\}$.  As far as a nonlinear polarization is
not present, the coefficients $a_n$ are time-independent. Conversely,
in the general case, taking for ${\bf P}_{NL}$ a standard third order
expansion, one has for the non-linear interaction Hamiltonian
\begin{eqnarray}
{\cal H}&=&-\langle{\int \epsilon_0 {\bf E}\cdot {\bf P}_{NL} dV
}\rangle
\nonumber
\\
&=&- \hspace*{-.5cm}\sum_{\omega_j+\omega_k
=\omega_l+\omega_m} \hspace*{-0.5cm}\Re\left[G_{jklm}~
a_j a_k a_l^* a_m^*\right]
\label{Ha1}
\end{eqnarray}
where $\langle \ldots \rangle$ is the time average over an optical cycle and 
the sum ranges over all distinct $4$-ples for which the condition
\BEQ
\omega_j+\omega_k
=\omega_l+\omega_m
\label{f:MLresonance}
\EEQ
holds, with $j,k,l,m=1,\ldots,N$.
The effective interaction  occurring among mode-amplitudes reads:
\begin{eqnarray}
G_{jklm}&=&\frac{\imath}{2}\sqrt{\omega_j\omega_k\omega_l\omega_m}
\\
\nonumber
&&\times
\int_V d^3r~
\chi^{(3)}_{\alpha\beta\gamma\delta}(\omega_j,\omega_k,\omega_l,\omega_m;
\mathbf r) 
\\
\nonumber
&& \hspace*{1cm}
\times ~ E^\alpha_j(\mathbf r) E_k^\beta(\mathbf r) E_l^\gamma(\mathbf r)
 E_m^\delta(\mathbf r)
\label{f:G4}
\end{eqnarray}
with $\alpha, \beta, \gamma, \delta = x,y,z$.  This coefficient
represents the spatial overlap of the electromagnetic fields of the
modes modulated by the non-linear susceptibility $\chi^{(3)}$.  The
disorder is induced, e.g., by the random spatial distribution of the
scatterers (as in random lasers) that leads to randomly distributed
modes and, hence, to random susceptibilities and 
couplings among quadruple of modes.
\\ \indent If the cavity is open, the mode set is no more complete,
the modes whose profile decays exponentially out 
of the cavity are taken for the
expansion (\ref{modeexpansion}), all the others form the radiation
modes.  Under standard approach \cite{HausBook, SakodaBook,
Hackenbroich:01, Hackenbroich03, Angelani06} the coefficients in the
expansion that weight the radiation modes can be expressed in terms of
the disordered cavity one, and this results into linear terms in the
Hamiltonian (open cavity regime). Thus, for an open cavity, Eq. (\ref{Ha1})
becomes
\begin{equation}
{\cal H}=-\Re\Biggl[~
\sum_{j<k} G^{(2)}_{jk} ~ a_j a_k^* +\hspace*{-.5cm}
\sum_{\omega_j+\omega_k=\omega_l+\omega_m}\hspace*{-.5cm}
G^{(4)}_{jklm} ~ a_j a_k a_l^* a_m^*\Biggr]
\label{Ha2}
\end{equation}
%% with 
%% \begin{equation}
%% G_{jk}=\frac{\imath}{2}\sqrt{\omega_j\omega_k}
%% \int_V d^3r~\chi^{(1)}_{\alpha\beta}(\omega_j,\omega_k;\mathbf r)~
%% E_j^\alpha(\mathbf r) E_k^\beta(\mathbf r)
%% \label{f:G2}
%% \end{equation}
%% {\bf riferimento a quantum physics, cita i lavori su quantum spin
%% glass di Theo...}\cite{Nieuwenhuizen95}

The Hamiltonian expressions, Eqs. (\ref{Ha1}), (\ref{Ha2}), can be  also
obtained starting from the corresponding Langevin dynamical equations,
as detailed, e.g., in Ref. [\onlinecite{Angelani06long}]:
\begin{eqnarray}
 \dot a_n(t)&=&\sum_j G^{(2)}_{jn} a_j
+\hspace*{-.5cm}\sum_{\omega_j+\omega_n=\omega_k+\omega_l} \hspace*{-.5cm}
G^{(4)}_{jkln} a_j^* a_k a_l + \eta_n(t)
\nonumber
\\
&=&-\frac{\partial \cal H}{\partial a_n^*}+ \eta_n(t)
\label{f:Langevin}
\end{eqnarray}
where $\eta_n(t)$ is a white noise, for which
\begin{equation}
\langle
\eta_j(t)\eta_k(t')\rangle = 2T\delta_{jk}\delta(t-t')
\label{f:noise}
\end{equation}
\\ \indent Here $T$ is a ``heat-bath'' temperature, whose physical
interpretation depends on the specific system.  In the case of a
random laser it represents the spontaneous emission and $k_B T\cong
\hbar/\tau$, with $\tau$ the amplifying level
lifetime. \cite{Angelani06long,YarivBook}
\\
\indent
Comparing Eq. (\ref{f:Langevin}) with the master equation for
mode-locking lasers in ordered cavities \cite{Haus00, HausBook}
\begin{eqnarray}
&& \dot a_n(t)\displaystyle= \left(g_n-\ell_n+i D_n\right)a_n
\nonumber
\\
&&\hspace*{1cm}+(\gamma-i\delta)
\hspace*{-.5cm}\sum_{\omega_j+\omega_n=\omega_k+\omega_l}\hspace*{-.5cm}
a_j^* a_k a_l + \eta_n(t)
\label{f:haus_master}
\end{eqnarray}
we can understand the physical role played by the parameters of the
probability distribution of the $G$'s.  Indeed, $g_n$ is the gain
coefficient of the $n$-th mode in a round-trip through the cavity,
$\ell_n$ the loss term, $D_n$ the group velocity of the wave packet,
$\gamma$ the coefficient of the saturable absorber (responsible for
passive mode-locking) and $\delta$ the coefficient of the Kerr lens
effect.  Neglecting the latter we can see that a system with positive
average of the $G_{jklm}$ corresponds to the presence of a saturable
absorber. In the case of peaked probability distribution for the
couplings $P(G)$, i.e., weak disorder, the system will tend to display
the same spectrum of many equally spaced modes typical of mode-locking
lasers. In the present formalism this will be a ferromagnetic
phase.  One might, then, wonder what happens when the disorder is
so strong to prevent the occurrence of this phase and, even, when the
random coefficient corresponding of $\gamma$ is negative (i.e., when
passive mode-locking is absent).  We will discuss this issue in
Sec. \ref{ss:satabs}.
\\
\indent
In the ``strong cavity limit'', the linear coupling between modes is 
negligible and $G_{mn}^{(2)}$ is diagonal (i.e., one accounts only for the
finite-life time of the modes) and
\begin{equation}
{\cal H}=-\Re\Biggl[
  \sum_{i=1}^N G^{(2)}_{ii} |a_i|^2 +\hspace*{-.5cm}
\sum_{\omega_{i_1}\!+\omega_{i_3}\!=\omega_{i_2}\!+\omega_{i_4}}\hspace*{-.5cm}
G^{(4)}_{i_1i_2i_3i_4}~ a_{i_1} a_{i_3} a_{i_2}^* a_{i_4}^*\Biggr]\text{.}
\label{Ha3}
\end{equation}

Note that the modes in the disordered cavity may display a different
degree of localizations, as in the case of disordered PhC.
Correspondingly, the distribution of the overlaps $G$ spreads.
Moreover, the constituents of the overlap integral are also very
difficult to calculate from first principles. Indeed, to our
knowledge, the only specific form of the non-linear susceptibility has
been computed by Lamb \cite{Lamb64} for a two-level system (without
disorder).  Eventually, to estimate the coupling distribution from the
experimental data is a very complicated inverse statistical
problem, cf., e.g., Refs.  [\onlinecite{Weigt09,Mora10}] and references
therein, and, so far, the reconstruction of the $G$'s, for example, from
measurements of random laser spectra has never been achieved.
 The interplay between susceptibility and spatial
distribution of modes leading to $G$'s is, then, a very challenging problem
that deserves a systematic and sophisticated treatment that goes
beyond the aim of the present work. 
\\ \indent In the following we will consider a mean-field approach in
which all modes are connected among each other.
We will, thus, approach the study of our model by means of Gaussian
distributed $G$'s with non-vanishing average, as detailed below.
\\
\indent
The leading regime considered in this work is, actually, driven by a {\it
quenched amplitude approximation}, which is obtained by retaining the
amplitudes $A_n=|a_n|$ (and correspondingly the energies of the modes)
as slowly varying w.r.t. the phase $\phi_n=\arg(a_n)$, such that the
resulting interaction Hamiltonian (retaining only those terms
depending on the phases, and considering the strong or closed cavity
regime, cf. Eq. (\ref{Ha3})), turns out to be \cite{Angelani06long, Leuzzi09}
\begin{eqnarray}
{\cal H}&=&-\hspace*{-.6cm }
\sum^{1,N}_{\omega_{i_1}\!+\omega_{i_3}\!=\omega_{i_2}\!+\omega_{i_4}}\hspace*{-0.9cm}'
\hspace*{.2cm} G_{i_1i_2i_3i_4} ~A_{i_1} A_{i_2}
A_{i_3} A_{i_4} \times
\label{f:Haphi2}
\\
\nn
&&\qquad\qquad\cos(\phi_{i_1}+\phi_{i_3}-\phi_{i_2}-\phi_{i_4})
\end{eqnarray}
where the sum $\sum '$ is limited to those terms that depend on the
phases (i.e., we neglect terms whose indices are such that the argument of
the cosine vanishes, e.g., $i_1=i_2$ and $i_3=i_4$)
and $G$ is assumed real-valued.
\\
\indent 
Actually, in the physical systems of our interest, 
it is not necessary that the resonant condition Eq. (\ref{f:MLresonance})
%$\omega_j+\omega_m=\omega_k+\omega_l$ 
for having four modes interact in the mode-locking regime is satisfied
exactly.  Indeed, it is enough that the mode combination tone
$\omega_{i_1}$ lies inside an interval around
$\omega_{i_2}+\omega_{i_4}-\omega_{i_3}$ corresponding to its
linewidth. \cite{MeystreBook} In the case, e.g., of the random laser,
in which many modes oscillate in a relative small bandwidth and are
densely packed in frequency space so that the their linewidth overlap,
this observation supports the further mean-field-like approximation
$\omega_i\simeq\omega_0$, $\forall ~ i$. In our model, therefore, the
spectral distribution of the angular frequencies will be considered as
strongly peaked around $\omega_0$ and $\omega_{i_1}+\omega_{i_3}\simeq
\omega_{i_2}+\omega_{i_4} \simeq 2\omega_0$ so that the ``selection
rule'' Eq. (\ref{f:MLresonance}) is always satisfied.
\\ \indent A suitable normalization and the introduction of an inverse
temperature-like parameter $\beta$ leads, eventually, from
Eq. (\ref{f:Haphi2}) to
\begin{equation}
\beta{\cal H}_J[\phi]=
-\beta\hspace*{-.5cm}\sum_{\footnotesize{\begin{array}{c} i_1<\!i_2, i_3<\!i_4
\\ 
i_1<\!i_3 \end{array}}}^{N}\hspace*{-.5cm}J_{\bf{i}}
\cos(\phi_{i_1}+\phi_{i_2}-\phi_{i_3}-\phi_{i_4})
\label{phase:Ham}
\end{equation}
with 
\begin{eqnarray}
&&\beta \equiv \frac{\langle A^2\rangle^2 }{k_B T_{\rm bath}}
\label{def:beta}
\\
&&J_{\bf{i}}=J_{i_1 i_2 i_3 i_4}\equiv\frac{G_{i_1 i_2 i_3 i_4}}{V^2}
\frac{A_{i_1}A_{i_2}A_{i_3}A_{i_4}}{\langle A^2\rangle^2}
\label{def:Ji}
\end{eqnarray}
where $T_{\rm bath}$ is the heat-bath temperature, variance of the
white noise $\eta$, cf. Eq. (\ref{f:noise}) induced by spontaneous
emission, and the squared volume factor guarantees thermodynamic
convergence ($\beta {\cal H} \propto V\sim N$).  The average energy
per mode is ${\cal E}_0 = \omega_0 \langle A^2\rangle$.  This is
proportional to the so-called {\em pumping rate} ${\cal P}$ induced on the
random laser by the pumping laser source.  We will define it as:
\begin{equation}
{\cal P}^2\equiv  J_0\frac{\langle
A^2\rangle^2}{k_B T_{\rm bath}}
\label{def:Pump}
\end{equation}
encoding the experimental evidence that decreasing the heat bath
 temperature \cite{Wiersma01} or increasing the energy of the pumping
 light source \cite{Leonetti10} has the same qualitative
 effect.  The proportionality factor $J_0$ in Eq. (\ref{def:Pump}) is
 a material dependent parameter function of the angular frequency
 $\omega_0$ of the peak of the average spectrum, cf. Eq. (\ref{f:G4}),
\begin{equation}
J_0=V\omega_0^2 \int_V ~d^3r~\chi^{(3)}(\omega_0; \bm r)
~|E(\bm r)|^2
\label{def:J0}
\EEQ in which $|E(\bm r)|= \langle E_n^2\rangle \sim 1/V$. Assuming 
that the non-linear susceptibility does not scale with the number of modes,
 the above integral
scales as $1/V$ and  $J_0$ does not scale with the size of the system.
The average of $J_{\bm i}$, instead, scales as $1/V^3$, according to
the definitions Eqs. (\ref{f:G4}) and (\ref{def:Ji}). 
\\ \indent To the sake of qualitative comparison with the outcome of
experiments the statistical mechanic inverse temperature $\beta$ can
be expressed in terms of the squared pumping rate as:
\begin{equation}
\beta = \frac{{\cal P}^2}{{J}_0}
\label{f:beta_pump}
\end{equation}

\subsection{Finite temperature Bose-Einstein condensates}
A similar situation is found in the finite temperature Bose Einstein
condensation with random potential.  The zero temperature
Gross-Pitaevskii equations \cite{Dalfovo99} reads as
\begin{equation}
\imath\hbar \frac{\partial \Phi}
{\partial t}=-\frac{\hbar^2}{2m}\nabla^2
\Phi+ V_{\text{ext}}(\bm r) \Phi+g |\Phi|^2 \Phi
\label{GP1}
\end{equation}
where $V_{\text{ext}}(\bm r)$ is an externally set disordered
potential and $g=4\pi \ell \hbar^2/m$, with $\ell$ being the $s$-wave
scattering length.  An analogous model holds for
reduced-dimensionality cases.
\\
\indent
The modes satisfy the time-independent linear Schroedinger
equation 
\begin{equation}
-\frac{\hbar^2}{2m}\nabla^2 \Phi_n+ V_{\text{ext}}(\bm r) \Phi_n=E_n \Phi_n
\end{equation}
Their interaction can be treated variationally by letting
 \BEQ
\Phi(\bm r,t)=\sum_n a_n(t) \Phi_n(\bfr) \exp{\left(-\imath
\frac{E_n}{\hbar}t\right)} \text{.}  \EEQ
\\
\indent
A finite temperature model for BEC is the Stoof equation,
\cite{Stoof99,Stoof01} which is here written as
\begin{eqnarray}
\label{Stoof}
\imath\hbar \frac{\partial \Phi}{\partial t}&=&
\left[1+\hbar\frac{\beta_K}{4}\Sigma^{K}(\bm r,t)\right]
\\
&&\times
\left[-\frac{\hbar^2}{2m}\nabla^2 \Phi+ V_{\text{ext}}(\bm r) \Phi+g
|\Phi|^2 \Phi\right]+\eta(\bm r,t)
\nn
\end{eqnarray}
with $\beta_K = 1/k_B T$ ($k_B$ is the Boltzmann constant) and where
the finite temperature noise is such that
\begin{equation}
\langle \eta^*(\bm r',t') \eta(\bm r,t) \rangle=\frac{\imath\hbar}{2}
\Sigma^{K}(\bm r,t)\delta
(t-t')\delta^{(3)}(\bm r-\bm r')\text{.}
\end{equation}
$\Sigma^{K}(\bm r,t)$ being the Keldish self-energy, which 
is imaginary valued (for its expression see
 Ref. [\onlinecite{Stoof01}]) and $\hbar \Sigma^K\propto -\imath
 \beta_K^{-2}$ (see Ref. [\onlinecite{Stoof04}]).  Expanding over
 the complete set of the zero temperature equations, one obtains
\begin{eqnarray}
&&\imath \hbar \dot a_n(t)=
-\imath \sum_j \alpha_{jn} a_j E_j 
e^{-\frac{\imath t}{\hbar}(E_j-E_n)}\\
\nn
&&\quad+\sum_{jkl} \left( G_{jkln}-\imath K_{jkln} \right) a_l^* a_j a_k 
e^{-\frac{\imath t}{\hbar}(E_j+E_k-E_l-E_n)}\\
\nn
&&\quad +\eta_n(t)
\end{eqnarray}
where $\eta_n(t) = \int d^3 {\bf r}~ \eta({\bf r},t) \phi_n({\bf r},t)$,
and the mode-overlap coefficients are defined as: 
\begin{equation}
G_{jklm}=g \int \Phi_j \Phi_k \Phi_l \Phi_m d^3{\bf r}
\end{equation}
and
\begin{equation}
K_{jklm}=\frac{\imath \beta_K \hbar g}{4} \int \Sigma^K(\bm r) \Phi_j
\Phi_k \Phi_l \Phi_m d^3{\bf r}\text{.}
\end{equation}
Finally, the linear
coupling coefficients come out to be
\begin{equation}
\alpha_{jk}=\frac{\imath \beta_K \hbar }{4} \int \Sigma^K(\bm r)
\Phi_j \Phi_k d^3{\bf r}
\end{equation}

While retaining the synchronous terms (such that $E_j+E_k-E_l-E_n=0$),
the resulting equations are, hence, of the same form of those reported
in Sec. \ref{ss:rl} for the disordered electromagnetic cavity, being
the energy of the eigenstates in place of the angular frequency.
Indeed, a strong coupling regime is attained when there is an enhanced
region for the density of states. Conversely, in other spectral regions,
both the linear and the nonlinear coupling terms are averaged out by the
rapidly oscillating exponential tails.
\\ \indent Let us consider, for example, a periodic external potential
with some degree of disorder. In this case, a Lifshitz tail
\cite{lifshitz64} is present, that is, a region with energies inside
the forbidden gap corresponding to localized modes.  This modes will
all have approximately the same energy $E\cong E_B$ where $E_B$ is the
band-edge energy, and will couple both among each other and with the
delocalized Bloch modes at the band-edge.  Correspondingly, the
relevant equations for the strongly coupled modes are
\begin{eqnarray}
\imath \hbar \dot a_n(t)&=&-\imath \sum_j \alpha_{jn} 
a_j E_B
\label{dynamicBEC}
\\ 
&&
\nn
+\sum_{jkl} \left( G_{jkln}-\imath
K_{jkln} \right) a_l^* a_j a_k+\eta_n(t)
\end{eqnarray}
The other modes (those far from the spectral gap) will be those mediating the
thermal bath.  The quenched amplitude approximation eventually leads to the
phase-dependent Hamiltonian, Eq.  (\ref{phase:Ham}).
\\ \indent As discussed in the following section of the manuscript,
even in the zero temperature limit a transition is expected. This
corresponds to the existence of a replica symmetry breaking transition
in Bose Einstein condensates for finite and vanishing temperature,
mediated by the degree of disorder and heuristically following the
phase diagram reported in Fig. \ref{fig:PhDi_P_R} below.

\subsection{Nonlinear optical propagation 
in disordered media and the zero temperature limit}

 The nonlinear optical propagation of a light beam is described by the
paraxial equation 
\BEQ \imath\frac{\partial A}{\partial z}+\frac{1}{2
k}\nabla_{x,y}^2 A+\frac{\Delta n}{2 k n}A=0 \EEQ
where $A$ is the optical amplitude, $k$ the wavenumber, $n$ is the
 bulk refractive index and $\Delta n$ is its perturbation due to
 disorder and optical nonlinearity (Kerr effect):
\BEQ
\frac{\Delta n}{2 k n}=U(x,y)+n_2 |A|^2. \EEQ 
The nonlinear coefficient $n_2$ can be either
positive (focusing) or negative (defocusing), while $U(x,y)$ can be a
perturbed (by disorder) periodical potential or a completely
disordered (speckle pattern) external potential.  The resulting
equation reads as 
\BEQ \imath\frac{\partial A}{\partial z}+\frac{1}{2
k}\nabla^2 A+ U(x,y) A+\frac{n_2}{2 k n}|A|^2 A=0\text{.}  \EEQ 
This formally corresponds (with different meanings for the variable) to the zero-temperature two-dimensional limit of the
Gross-Pitaevskii equations detailed above, cf. Eq. (\ref{GP1}).
\\ \indent In this case, as well, the field can be expanded in terms of
transversely localized (in two dimensions they are always localized)
electromagnetic modes, the energies being replaced by their
propagation wave-vectors. When there are bunch of modes such that
their wave-vectors are similar, these will be strongly coupled and
result into dynamical equations like Eqs. (\ref{f:Langevin}),
(\ref{dynamicBEC}).  This approach can be extended to
three-dimensional propagation, encompassing the dynamics of
ultra-short pulses in random media as will be reported elsewhere.
\\ \indent The replica symmetry breaking transitions investigated in
the following will in general correspond to varying coherence
properties of the beam, eventually resulting in unstable speckle
patterns.  The $\beta\to\infty $ limit of the statistical mechanical
formulation of the problem has to be taken in this case (see, e.g.,
Ref. [\onlinecite{Leuzzi01}] for a simple case example in the
framework of constraint satisfaction problems).

\section{Randomness in mode-coupling coefficients}
\label{sec:random}

Let us consider our model Hamiltonian, Eq. (\ref{f:Haphi2}), in the
mean-field fully connected approximation in which the
non-vanishing components of the 
 four index
tensor $J_{i_1,i_2,i_3,i_4}=J_{\bf i}$ are distributed as 
\BEA {\overline{J_{\bf
i}}}&=&J_0/N^3
\label{f:Jave}
\\
{\overline
{(J_{\bf i}-{\overline{J_{\bf i}}})^2}}&=&\sigma^2_J/N^3
\label{f:Jvar}
\EEA
The coefficient $J_0$ was already introduced in the case of random lasers,
cf. Eq. (\ref{def:J0}), and $N$ is the number of dynamic variables (mode
phases) of the system, proportional to the volume $V$.  The overbar
denotes the average over the disorder.
\\ \indent To quantify the amount of disorder, we introduce the
``degree of disorder'' parameter, i.e., a size independent ratio
between the standard deviation of the distribution of the coupling
coefficients $J_{\bf i}$ and their mean:
\BEQ R_J\equiv \frac{\sigma_J}{J_0}
\EEQ
 The limits $R_J\rightarrow 0$ and $R_J\rightarrow\infty$ correspond,
respectively, to the completely ordered and  disordered case.
The other relevant parameter for our investigation is the inverse
temperature $\beta$.  For random lasers it is
related to the normalized pumping threshold for ML, defined in our
model as, cf. Eq. (\ref{f:beta_pump}),
 \footnote{If $J_0=0$ we are in the completely disordered case
$R_J=\infty$ (also realizable by means of a finite $J_0$ and
$\sigma_J^2=\infty$). In Ref. [\onlinecite{Angelani06long}] ${\cal P}$
has been defined as $\sqrt{\beta k_B T_{\rm bath}}$, simply amounting
to an adimensional rescaling ${\cal P}\to {\cal P}\sqrt{J_0\beta_{\rm
bath}}$ w.r.t. our model case.}
\BEQ
{\cal P}=\sqrt{\beta J_0}=\sqrt{\frac{\bar \beta}{R_J}}
\EEQ
where $\bar\beta\equiv\beta\sigma_J$. \cite{Leuzzi09} In general,
$\beta$ increases as the strength of nonlinearity increases or the
amount of noise is reduced.

\subsection{The ordered limit, saturable absorbers 
in random lasers, defocusing versus focusing}
\label{ss:satabs}
With specific reference to the laser systems, as $J_0$ grows the
effect of disorder is moderated and for small enough $R_J$ the model
corresponds to the ordered case, previously detailed in Ref.
[\onlinecite{Angelani07}].  As also previously reported in
Ref. [\onlinecite{Gordon03}], a passive mode-locking (PML) transition
is predicted as a paramagnetic/ferromagnetic transition
occurs in $\beta$.
\\ \indent Indeed, in our units, when $R_J\rightarrow0$, ${\cal
P}={\cal P}_{\rm PML}\cong 3.819$ (see Fig. \ref{fig:PhDi_P_R}), in
agreement with the ordered case.  \cite{Angelani06} ${\phantom
.}$\footnote{A factor of $8$ has to be considered because of the
over-counting of terms in the Hamiltonian of the model studied in
Ref. [\onlinecite{Angelani06}] with respect to
Eq. (\ref{f:Haphi2}). This factor can be absorbed into the temperature
yielding the pumping threshold ${\cal P}_{\rm PML}=\sqrt{8/T_0}$.  If
we insert $T_0\cong 0.717$, i.e., the temperature at which the FM
phase first appears in complete absence of disorder we obtain ${\cal
P}_{\rm PML}\cong 3.34$. This also exactly corresponds to the spinodal
value of ${\cal P}=3.3412$ for $J_0/\sigma_J\to\infty$ in the present
model.} As explained below, the deviation from this value quantifies
an increase of the standard ML threshold ${\cal P}_{\rm PML}$ due to
disorder.  The specific value for ${\cal P}_{\rm PML}$ will depend on
the class of lasers under consideration (e.g., a fiber loop laser or a
random laser with paint pigments), but the trend of the passive ML
threshold with the strength of disorder $R_J$ in
Fig. \ref{fig:PhDi_P_R} has a universal character.  The pumping rate
$\mathcal{P}$ contains $J_0$: for a fixed disorder the threshold will
depend on the nonlinear mode-coupling.
\\ \indent A key point here is that the transition from continuous
wave to passive mode-locking (PM $\to$ FM) only occurs for a specific
sign of the mean value of the coupling coefficient $J_0$, as shown in
Fig.  (\ref{fig:PhDi_T_J0}).  Comparing Eqs. (\ref{f:Langevin}) and
(\ref{f:haus_master}) one observes that this formally corresponds to
the presence of a saturable absorber in the cavity (see also
Ref. [\onlinecite{Haus00}] and Sec. \ref{ss:rl}).
In typical random lasers such a device is not present, and, hence,
this ferromagnetic transition is not expected. 
\\
\indent
On the other hand, the
reported phase diagram, Fig. (\ref{fig:PhDi_T_J0}) predicts that
starting from a standard laser supporting passive/mode-locking and
increasing the disorder the second order transition acquires the
character of a glass transition.
A notable issue is that this phase-locking transition (normally ruled
out for ordered lasers without a saturable transition), spontaneously
occurs increasing $\beta$, as an effect of the disorder and the
resulting frustration.
\\ \indent With reference to nonlinear waves, the spontaneous
phase-locking process is expected for a specific sign of the nonlinear
susceptibility (corresponding to repulsive interactions for BEC and
defocusing nonlinearities for optical spatial beams), for $T=0$,
amounting to $J_0/\sigma_J>0$ in Fig. (\ref{fig:PhDi_T_J0}) (the
threshold is at $J_0/\sigma_J\cong 4$). For example, for a
nonlinear optical beam propagating in a disordered medium, it is
expected that above a certain degree of disorder, there is a transition
from a coherent regime to a ``glassy coherent phase'', characterized
by a strong variation from shot to shot of the speckle pattern and,
more in general, of the degree of spatial coherence.

\section{Fundamentals of Statistical Mechanics of Disordered Systems}
\label{sec:statmech}
Hereby we report an extremely concise summary of ideas and techniques
developed to deal with disordered systems. The aim is to let the
non-expert reader find his/her way through the computation of the
properties of our model that we present in Sec. \ref{sec:computation}
and App. \ref{app}.
\subsubsection{Disorder and frustration:
\\ quenched disorder as technical tool.}
\label{sec:frustration}
The main issue determining complex features, not present in ordered
systems and involving collective processes that cannot be understood
just looking at local properties, is {\em frustration}. This is
usually a the consequence of disorder, not necessarily {\em quenched }
disorder, though.  Indeed, also in materials whose effective
statistical mechanic representation is carried out through
deterministic potentials (as, e.g., for colloidal particles), a
geometry-induced disorder can set up, determining frustration and a
consequent multitude of degenerate stable and metastable states
typical of glasses
\cite{Barrat90,Hansen95,Kob94,Kob95a,Kob95b,Sciortino99,Mezard99,Coluzzi00}
and spin-glasses. \cite{Marinari94a,Marinari94b,Cugliandolo95}
Quenched disorder, i.e., the explicit appearance of random
coefficients in the Hamiltonian, allows an analytic computation, but
the results are general and do not depend on the specific source of
frustration.

\subsubsection{Statistical mechanics of a disordered system: 
\\ the replica trick.}  In the presence of quenched disorder, one can
 compute the statistical mechanics of the system, averaging
 over the probability distribution of the disorder. In order to do
 this the so-called replica trick
 \cite{Sherrington75,Parisi79,Parisi80,MPVBook} can be adopted, or,
 else, the equivalent cavity method.  \cite{Mezard86,MPVBook}
\\
\indent
The free energy of a single
disordered system sample, denoted by $J$, is 
$\Phi_J=-1/\beta \log Z_J$.
Correspondingly, the physically relevant average free energy
can be written as 
\BEQ \Phi=-\frac{1}{\beta} {\overline {\log Z_J}}
\label{f:Fen_ave}
\EEQ
 where the overbar denotes the
average over the distribution of the $J$'s. The latter coincides with
the thermodynamic limit of any $\Phi_J$ according to the
self-averaging property required in order to have macroscopic
reproducibility of experiments  (the thermodynamics of a huge system
does not depend on the local distribution of interaction
couplings).
\\ \indent To perform the average in Eq. (\ref{f:Fen_ave}) is highly
non trivial and one can proceed by considering $n$ copies of the
system, Eq. (\ref{f:Haphi2}),
\begin{equation}
{\cal H}[\{\phi\}] \to \sum_{a=1}^n {\cal H}[\{\phi^{(a)}\}]
\label{f:Hphirep}
\end{equation}
The average free energy per spin can, then, be
computed in the replicated system, as
\begin{eqnarray}
&&\beta \Phi=-\lim_{N\to \infty}
\frac{1}{N}~{\overline {\log Z_J}}=-\lim_{N\to \infty}
\lim_{n\to
0}\frac{{\overline {Z^n_J}}-1}{N~n} 
\nonumber
\\
\label{f:Phi_rep}
\end{eqnarray}  
where the average of the generic power of the partition function
${\overline {Z^n_J}}$ is somehow computed for a finite integer $n$
and, eventually, the analytic continuation to real $n$ and the limit
$n\to 0$ are performed.

\subsubsection{Oddities of the replica formulation.}
Actually, to evaluate ${\overline {Z^n_J}}$, one makes use of the
saddle point approximation holding for large $N$ (see Appendix
\ref{app} for the specific case considered in this work). That is, one
practically inverts the limits $N\to \infty$ and $n\to 0$ as expressed
in Eq. (\ref{f:Phi_rep}). Yet, the method works.  It took many years
to rigorously overcome this oddity and a mathematical proof of the
existence of the free energy can be found in
Refs. [\onlinecite{Guerra03,Talagrand06}].

\subsubsection{A probability distribution as an order parameter.}
The main novelty of the characterization of the spin-glass phase,
historically first obtained by the replica method and subsequently
confirmed by other methods, is that the order parameter is a whole
probability distribution function describing how different
thermodynamic states are correlated. The degree of the correlation
between two states is called {\em overlap}. In mean-field theory
different states exist that can be more or less correlated according
to their distance on a tree-like hierarchical space called {\em
ultrametric}. \cite{Mezard84}

\subsubsection{ Complexity as a well-defined thermodynamic potential.}
Besides numerous and hierarchically organized globally stable states,
glasses also display a large number of metastable states, that is,
excited states of relatively long lifetime. In the mean-field theory
such lifetime is, actually, infinite in the thermodynamic limit
because of the divergence of the free energy barriers with the size of
the system, see, e.g., Ref.  [\onlinecite{CC05}].  This means that,
contrarily to what happens in real glasses, \cite{Leuzzi07} the number
of metastable states at a given observation timescale does not change
with time (after a given transient period).  Below a certain
temperature (called {\em dynamic} or {\em mode coupling} temperature),
the number ${\cal N}$ of metastable states grows exponentially with
the size $N$ of the system ($N$ being the number of modes in our
cases).  One can then define an entropy-like function counting the
metastable states as
\begin{equation}
\Sigma \equiv \frac{1}{N}\log {\cal N} \ .
\label{def:complexity}
\end{equation}
This is called {\em configurational entropy} in the framework of
structural glasses, else {\em complexity} in spin-glass theory and its
applications to constraint satisfaction and optimization problems.
One can further look at the metastable states of equal free energy
density $f$: ${\cal N}(f)=\exp N\Sigma(f)$ and at the free energy
interval, above the equilibrium free energy $f_{\rm eq}$, 
in which the complexity
is non zero: $f\in [f_{\rm eq}:f_{\star}]$.

\section{Statistical mechanical properties }
\label{sec:computation}
Starting from the Hamiltonian, Eq. (\ref{f:Haphi2}), replicated
according to the prescription Eq. (\ref{f:Hphirep}), and averaging
over the disorder with the Gaussian probability expressed by Eqs. 
(\ref{f:Jave})-(\ref{f:Jvar}), one obtains the
following expression for the average of the $n$-th power of the partition
function, cf.  Appendix \ref{app}:
\BEA 
&&{\overline {Z_J^n}}=\int{\cal
D}\mathbf{Q}~{\cal D}\mathbf{\Lambda}~ e^{-N~n~G\left[
\mathbf{Q},\mathbf{\Lambda}\right]}
\label{f:Zintegral}
\\ \nn &&n~G\left[
\mathbf{Q},\mathbf{\Lambda}\right]=n~A\left[
\mathbf{Q},\mathbf{\Lambda}\right] +\log Z_\phi\left[\mathbf{\Lambda}
\right] 
\\ && n~A \left[\mathbf{Q},\mathbf{\Lambda}\right] \equiv
-\frac{\beta^2\sigma_J^2}{32} \sum_{a=1}^n\Bigl(1+\left| \tilde
r_a\right|^4\Bigr) -\frac{\beta J_0}{8}\sum_{a=1}^n\left| \tilde
m_a\right|^4 \nn 
\\ \nn
&&\hspace*{.5cm}-\frac{\beta^2\sigma_J^2}{16}\sum_{a<b}^{1,n}\Bigl(q_{ab}^4+\left| r_{ab}\right|^4\Bigr)
-\sum_{a<b}^{1,n}\bigl[
q_{ab}\lambda_{ab}+\Re\left(r_{ab}\bar\mu_{ab}\right)\bigr] \\ &&\hspace*{1cm}-
\sum_{a=1}^n\Re\bigl[\tilde r_a{\bar{\tilde \mu}}_a+\tilde m_a\nu_a
\bigr]
\EEA 
\BEA
 &&Z_\phi\left[\mathbf{\Lambda} \right]\equiv
\int\prod_{a=1}^n d\phi_a e^{-\beta{\cal H}_{\rm
eff}[\{\phi\};\mathbf{\Lambda}]}
\label{f:Zphi_main}
\\
&&-\beta{\cal H}_{\rm eff}[\{\phi\};\mathbf{\Lambda}]\equiv 
\sum_{a<b}^{1,n}\Re\bigl[e^{\imath(\phi_a-\phi_b)}\lambda_{ab}
+e^{\imath(\phi_a+\phi_b)}\bar\mu_{ab}\bigr]
\nn
\\
\label{f:Hphi_main}
&&\hspace*{2cm} +\sum_{a=1}^n\Re\bigl[ e^{2\imath \phi_a}{\bar{\tilde
\mu}}_a+e^{\imath \phi_a}\bar\nu_a \bigr] 
\EEA
\BEA
\nn &&{\cal
D}\mathbf{Q}\equiv \prod_{a<b}^{1,n}N^2 dq_{ab} dr_{ab}\times
\prod_{a=1}^n N^2 d\tilde r_a d\tilde m_a \\ \nn &&{\cal
D}\mathbf{\Lambda}\equiv \prod_{a<b}^{1,n} \frac{d\lambda_{ab}}{2\pi}
\frac{d\mu_{ab}}{2\pi}\times \prod_{a=1}^n \frac{d\tilde
\mu_a}{2\pi}\frac{d\nu_a}{2\pi} \EEA 
where $\mathbf{Q}=\{q,r,\tilde
r,\tilde m\}$ and $\mathbf{\Lambda}=\{\lambda,\mu,\tilde\mu,\nu\}$.
The overlap matrices $q_{ab}$, $\lambda_{ab}$ are real-valued,
whereas the others have  complex elements.
\\
\indent
The
integral Eq. (\ref{f:Zintegral})
is evaluated by means of the saddle point approximation
(valid for large $N$).
The above expressions need a form of the matrices $q_{ab}$, $r_{ab}$,
$\lambda_{ab}$ and $\mu_{ab}$ to be completed. Contrarily to what
might seem reasonable, the form providing the thermodynamically stable
solution is not the one in which all replicas are equivalent, i.e.,
all elements in the matrices $q_{ab}$, $r_{ab}$, $ \lambda_{ab}$,
$\mu_{ab}$ are equal. One must, thus, resort to a spontaneous {\em
Replica Symmetry Breaking}.  In Appendix \ref{app} we report the
computation of thermodynamics both in the Replica Symmetric (RS)
approximation and in the ``one step'' Replica Symmetry Breaking Ansatz
(1RSB), i.e., the exact solution for the system under probe. In the
following we, thus, analyze the properties of the latter
solution.
\\
\indent
Spin-glass systems described by more-than-two-body interactions,
cf. Eq. (\ref{f:Haphi2}), are known to have low temperature phases that
are stable under the 1RSB Ansatz. \cite{Gardner85,Crisanti92}
\footnote{Since in our model the dynamic variables are continuous
phases, the whole low $T$ phase is consistently described by the 1RSB
solution, unlike models with discrete variables such as the Ising
$p$-spin model \cite{Gardner85} where a further transition occurs at
the so-called Gardner temperature.}  Under this Ansatz,
taking the $n\to 0$ limit, the free energy functional $\beta \Phi$
reads, cf. Appendix \ref{app},
\begin{eqnarray}
\label{f:repPhi_1rsb}
&&\beta \Phi(m;{\bf Q}_{\rm sp}^{(1)},{\bf \Lambda}_{\rm sp}^{(1)})=
G(m;{\bf Q}_{\rm sp}^{(1)},{\bf \Lambda}_{\rm sp}^{(1)})= 
\\
\nn
&&\quad=-\frac{\bar\beta R_J}{8}|{\tilde m}|^4
-\frac{{\bar\beta}^2}{32}\Bigl[
1
-(1-m)\left(q_1^4+|r_1|^4\right)
\\
\nonumber
&&\qquad -m\left(q_0^4+|r_0|^4\right)+|{r_d}|^2
\Bigr]
 -\Re\Big[\frac{1-m}{2}\left(\lambda_1q_1+\bar\mu_1r_1\right)
\\
\nonumber
&&\qquad+\frac{m}{2}\left(\lambda_0q_0+\bar\mu_0r_0\right)
-{\bar \mu_d} {r_d} - {\bar \nu}{\tilde m}\Bigr]+ \frac{\lambda_1}{2}
 \\
\nonumber
&&\qquad -\frac{1}{m}\int {\cal D}[\bm{0}]\log\int{\cal D}[\bm{1}]
\left[\int_0^{2\pi}\!d\phi~\exp{\cal L}(\phi; \bm{0},\bm{1})
\right]^m
\end{eqnarray}
where $\bm{0}=\{x_0,\zeta_0^R,\zeta_0^I\}$, 
$\bm{1}=\{x_1,\zeta_1^R,\zeta_1^I\}$, 
${\cal D}[\bm{a}]$ is the product of three Normal distributions
and
\begin{eqnarray}
\nonumber
&&{\cal L}(\phi;\bm{0},\bm{1})\equiv\Re\Bigl\{e^{\imath\phi}\Bigl[
\bar\zeta_1\sqrt{\Delta \lambda-|\Delta \mu|}+
\bar\zeta_0\sqrt{\lambda_0-|\mu_0|}+
\\
\label{f:calL}
&&\quad x_1\sqrt{2\Delta \bar\mu}
+x_0\sqrt{2\bar\mu_0}
+\bar\nu\Bigr]
+e^{2\imath\phi}\left(\bar \mu_d-\frac{\bar\mu_1}{2}\right)
\Bigr\}
\end{eqnarray}
with $\Delta\lambda=\lambda_1-\lambda_0$, $\Delta \mu=\mu_1-\mu_0$.
For later convenience
 we define the following averages over the action
$e^{\cal L}$, cf. Eq. (\ref{f:calL}): 
\BEA
c_{\cal L}\equiv \langle
\cos\phi\rangle_{\cal L} \equiv \frac{\int_0^{2\pi} d\phi~\cos\phi~e^{\cal L}} 
{\int_0^{2\pi} d\phi~e^{\cal L}} 
\\
s_{\cal L}\equiv \langle
\sin\phi\rangle_{\cal L}\equiv \frac{\int_0^{2\pi} d\phi~\sin\phi~e^{\cal L}} 
{\int_0^{2\pi} d\phi~e^{\cal L}} 
\EEA
The values of 
the order parameters 
$\lambda_{0,1}, \mu_{0,1}, \mu_d$ and $\nu$
are 
yielded by
\begin{eqnarray}
&&\lambda_{0,1}=\frac{{\bar\beta}^2}{4} \left(q_{0,1}\right)^3~
;\quad 
\mu_{0,1}=\frac{{\bar\beta}^2}{4}|r_{0,1}|^2~ r_{0,1}
\label{f:la_q_1rsb}
\\
&&\tilde\mu= \frac{{\bar\beta}^2}{8}|\tilde r|^2{\tilde r}~~
;\qquad \nu=\frac{\bar\beta R_J}{2}|\tilde m|^2{\tilde m}
\label{f:mu_r_1rsb}
\end{eqnarray}
The parameter $m$ (without tilde!), whose meaning will be discussed
 below, takes values in the interval $[0,1]$.  The remaining
 parameters are obtained by solving the self-consistency equations:
\begin{eqnarray}
\label{f:speq_q1_R}
&&\hspace*{-4mm}
q_1=\langle \langle c_{\cal L}^2 \rangle_{m}\rangle_{\bf 0}+
\langle \langle  s_{\cal L}^2\rangle_{m}\rangle_{\bf 0}
\\
\label{f:speq_q0_R}
&&\hspace*{-4mm}
q_0=\langle \langle c_{\cal L} \rangle_{m}^2\rangle_{\bf 0}+
\langle \langle  s_{\cal L}\rangle_{m}^2\rangle_{\bf 0}
\\
\label{f:speq_r1R}
&&\hspace*{-4mm}
r_1= \langle \langle c_{\cal L}^2 \rangle_{m}\rangle_{\bf 0}
-\langle \langle s_{\cal L}^2 \rangle_{m}\rangle_{\bf 0}
+2 \imath \langle \langle c_{\cal L} s_{\cal L} \rangle_{m}\rangle_{\bf 0}
\\
\label{f:speq_r0R}
&&\hspace*{-4mm}
r_0=\langle \langle c_{\cal L} \rangle_{m}^2\rangle_{\bf 0}
-\langle \langle s_{\cal L} \rangle_{m}^2\rangle_{\bf 0}
+2 \imath \langle \langle c_{\cal L} \rangle_{m}\rangle_{\bf 0}
\langle \langle  s_{\cal L} \rangle_{m}\rangle_{\bf 0}~~
\\
\label{f:speq_rd}
&&\hspace*{-4mm}
\tilde r=\langle \langle \langle
e^{2\imath \phi}
\rangle_{\cal L}\rangle_{m}\rangle_{\bf 0};
\qquad\tilde m=\langle \langle 
\langle e^{\imath \phi}
\rangle_{\cal L}\rangle_{m}\rangle_{\bf 0}
\end{eqnarray}
where the averages are defined as
\begin{eqnarray}
\langle(\ldots )\rangle_m&\equiv&
\frac{\int{\cal D}[\bm{1}](\ldots )\left[\int_0^{2\pi}\!
d\phi~e^{{\cal L}(\phi;\bm{0},\bm{1})}
\right]^m}
{\int{\cal D}[\bm{1}]\left[\int_0^{2\pi}\!d\phi~e^{{\cal L}
(\phi;\bm{0},\bm{1})}\right]^m}
\\
\langle(\ldots )\rangle_{\bf 0}
&\equiv&
\int{\cal D}[\bm{0}](\ldots )
\end{eqnarray}
These equation are solved numerically by an iterative method.
The overlap parameters $q_{0,1}$  are real-valued,
whereas $r_{0,1}, {\tilde r}$ and ${\tilde m}$ 
are complex. ``One step'' parameters $X_{0,1}$ ($X=q,r$)  enter with a
probability distribution that can be parametrized by the
 so-called
{\em replica symmetry breaking parameter} $m$, such that
\BEQ
P(X)=m~\delta(X-X_0)+(1-m)\delta(X-X_1).
\label{f:1RSB_prob}
\EEQ
The resulting independent parameters (there are ten of them) that can
be evaluated by solving Eqs. (\ref{f:speq_q1_R})-(\ref{f:speq_rd})
must be combined with a further equation for the parameter $m$.  This
is strictly linked to the expression for the {\em complexity} function
of the system.

\section{Complexity}
\label{sec:complexity}
In the order parameter Eqs. (\ref{f:la_q_1rsb})-(\ref{f:speq_rd}) $m$
 is left undetermined.  An additional condition is needed to fix the
 value for this parameter.  The first possibility is treating $m$ as a
 standard order parameter: in this case the thermodynamic state
 corresponds to extremizing the replicated free energy (thus, {\em
 maximizing} it
\footnote{Technically speaking, this is due to the fact that all the
terms of the free energy functional depending on two replicas
observables have $n-1$ or $n-m$ factors in front, cf. App. \ref{app},
and in the $n\to 0$ limit this factors change.}), i.e., implementing
the self-consistency equation
\BEQ \frac{\partial \Phi(m;{\bf Q}_{\rm sp},{\bf \Lambda}_{\rm
sp})}{\partial m}=0 
\label{f:dphi_dm}\EEQ
The highest temperature at which
a solution exists with $m\leq 1$ furnishes a transition temperature between
paramagnet and glassy phase: the Kauzmann or {\em static} temperature
($T_s$). This is an {\em equilibrium thermodynamic} phase transition.
\\ \indent This approach, however, does not reflect the known physical
circumstance that a glassy system exhibits excited metastable states
also at temperature $T$ above $T_s$, \footnote{The very existence of a
Kauzmann temperature, also called the {\em ideal glass transition
temperature}, in structural glasses is, actually, a matter of debate.}
where the equilibrium phase is paramagnetic.  Vitrification, indeed,
is due to the presence of a not vanishing complexity at a temperature
above $T_s$ (and below some $T_d>T_s$), i.e., to the presence of a
number of energetically equivalent states with free energy $f>f_{\rm
eq}(T)$.  Since, however, energy barriers tend to infinity in the
thermodynamic limit in the mean-field approximation, the system
dynamics is forever trapped in one of these states for $T<T_d$.
The temperature $T_d$ is, thus, called {\em dynamic} transition temperature.
\\ \indent Across this transition the complexity $\Sigma$, defined in
Eq. (\ref{def:complexity}), starts being different from zero. Exactly
at $T=T_d$ the complexity as a function of free energy, $\Sigma(f)$,
has a delta-shaped non-zero peak 
at the free energy $f_1$ which corresponds to a maximum
of $\Sigma(m)$ for a value of $m=m(f_1)=1$. In our 1RSB formalism:
\BEQ
\frac{\partial \Sigma(m;{\bf Q}_{\rm sp},{\bf \Lambda}_{\rm sp})}{\partial m}=0
\EEQ
As $T$ decreases ($T_s<T<T_d$), the complexity is not vanishing for an
increasing range of free energies $f^*>f>f_1$ that corresponds to a
range for $m$: $m^*< m <1$. The complexity shows a maximum for $m\leq
1$ at $m_*$ ($f=f_*$) solution of $d\Sigma/dm=0$, while it is at its
minimum value for $m=1$ and $f=f_1$. We stress that this is not a
solution to Eq. (\ref{f:dphi_dm}).
\\ \indent Lowering the temperature, at $T=T_s$ the minimum value of
complexity - corresponding to $m=1$ - vanishes, i.e., it is a solution
to Eq. (\ref{f:dphi_dm}), and $f_1=f_{\rm eq}$ corresponds to the free
energy density of the global glassy minima of the free energy
landscape: as mentioned above, we are in presence of a thermodynamic
phase transition and the thermodynamic stable phase is a glass.
\\ \indent The physically significant value for $m$ is $m^*$,
corresponding to the maximum of $\Sigma$. It denotes the value of
free energy $f^*$ where the number of states is maximum and
exponentially higher than the number of states at any $f<f^*$, and,
hence, the most probable (among those of the metastable states).  At
the thermodynamic transition point from the paramagnetic state to the glassy
($T=T_s$) it holds $f_{PM}=f_1=f_{\rm eq}=\Phi$. \footnote{
The paramagnetic phase exists as metastable also at $T<T_s$ but the
phase space is disconnected and the ergodicity is broken because of
infinite barriers.}
Below $T_s$ $f_1<f_{\rm eq}$ (hence, $\Sigma(f_*)<\Sigma(f_{\rm eq})=0$)
and the physically relevant $\Sigma(f)$ has a support $[f_{\rm eq},f_*]$.
\\ \indent In the following we will analyze the whole complexity
vs. free energy curve $\Sigma(f)$ at given $\beta,J_0$ and the
behavior of the minimal positive complexity $\Sigma(T)$ (and $\Sigma({\cal
P})$) between $T_s$ and $T_d$.

\subsection{Computing the complexity functional}

In Eq. (\ref{def:complexity}) one needs to know the number of
metastable states, that are the local minima of the free energy
landscape. Would we know the landscape, though, we would have solved
the problem already. If self-consistency equations for local order
parameters are known, a possible analytic approach to get information
on the complex landscape is to guess a trial free energy functional
whose stationary equations lead back to the self-consistency
equations. This is what Thouless, Anderson and Palmer (TAP) proposed
in the framework of spin-glasses starting from the self-consistency
equations for local magnetizations. \cite{TAP77} Starting from TAP
functional and TAP equations and considering solutions to the TAP
eqs. as {\em states} (with some assumptions to be {\em a posteriori}
satisfied) one can build the functional $\Sigma$ from
Eq. (\ref{def:complexity}), cf., e.g.,
Refs. [\onlinecite{Bray80,Crisanti03,Crisanti03b,Annibale03,
Crisanti04,Crisanti04b,Aspelmeier04,Crisanti05}].
\\
\indent 
A comparative study
to the TAP-derived complexity functional and the replicated free
energy, computed in a general scheme that includes the Parisi Ansatz,
\cite{Mueller06}  allows to show that the Legendre Transform of
$\Phi$ with respect to the single state free energy coincides with
Eq. (\ref{def:complexity}).
According to this approach,
in our model the complexity can, thus, be explicitly
computed as the Legendre transform of Eq. (\ref{f:repPhi_1rsb}):
\begin{eqnarray}
&&\Sigma(m;{\bf Q}_{\rm sp},{\bf \Lambda}_{\rm sp})
\label{f:Sigma}
\\
\nn
&&\hspace*{1cm}= \min_m \left[-\beta m \Phi(m)+\beta m f\right]
\\
\nn
&&\hspace*{1cm}=  \beta m^2\frac{\partial\Phi}{\partial m}
\\
&&\hspace*{1cm}=\frac{3}{4}\beta^2m^2
\left(|q_1|^4+|r_1|^4-|q_0|^4-|r_0|^4\right)
\nonumber
\\
\nonumber
& &\hspace*{1.1cm}+\int {\cal D}[\bm{0}]\log\int{\cal D}[\bm{1}]
\left[\int_0^{2\pi}\!\!\!\!d\phi~\exp{\cal L}(\phi; \bm{0},\bm{1})
\right]^m
\\
&&\hspace*{1.1cm} -m\int {\cal D}[\bm{0}]\langle\log
\int_0^{2\pi}\!\!\!\!d\phi~\exp{\cal L}(\phi; \bm{0},\bm{1})
\rangle_m
\nonumber
\end{eqnarray}
where the single state free energy 
\begin{equation}
f=\frac{\partial (m\Phi)}{\partial m}
\end{equation}
is
conjugated to $m$. Since the above expression is proportional to
$\partial \Phi/\partial m$, equating $\Sigma=0$ 
provides the missing equation to determine the order parameters
values.

\section{Phase Diagram and Complexity}
\label{sec:phdi}

By varying the normalized pumping rate ${\cal P}$ and the degree of
disorder $R_J$, we find three different phases, as shown in Fig.
\ref{fig:PhDi_P_R} in the $({\cal P}, R_J)$ plane and in Figs.
 \ref{fig:PhDi_T_J0}, \ref{fig:PhDi_T_J0_det} in the $(T,J_0)$ plane.

\begin{figure}
\includegraphics[height=.99\columnwidth, angle=270]{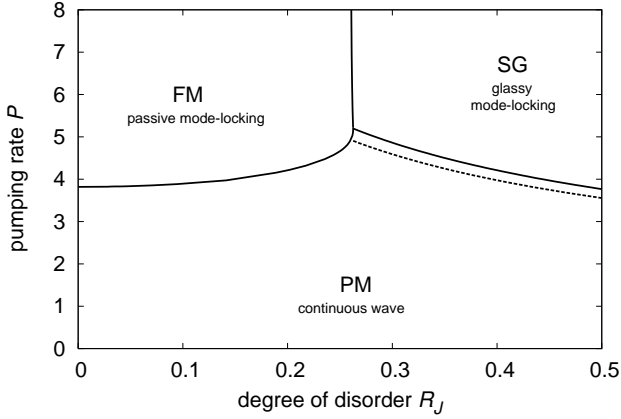}
\caption{Phase diagram in the ${\cal P}, R_J$ plane.  Three
phases are present: PM (low ${\cal P}$), FM (high ${\cal P}$/weak
disorder) and SG (high ${\cal P}$/strong disorder).  The full lines
are thermodynamic transitions, the dashed line represents the dynamic
PM/SG transition. }
\label{fig:PhDi_P_R}
\end{figure}
\begin{figure}
\includegraphics[height=.99\columnwidth, angle=270]{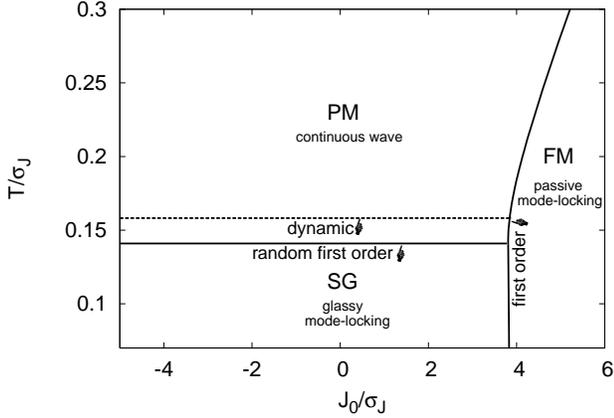}
\caption{Phase diagram in the plane $J_0,T$ in $\sigma_J$ units. Also
negative $J_0$ are considered. Three phases are found: PM (high $T$,
low $J_0$), FM (low $T$/large $J_0$) and SG (low $T$/low or negative
$J_0$).  The full lines are thermodynamic transitions: {\em random}
first order between PM and SG and {\em standard} first order between
PM and FM and between SG and FM. The dashed line represents the
dynamic PM/SG transition. }
 \label{fig:PhDi_T_J0}
\end{figure}
\begin{figure}
\includegraphics[height=.99\columnwidth, angle=270]{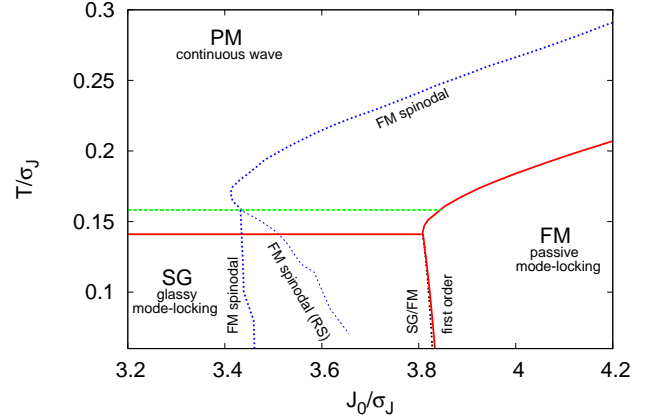}
\caption{(Color online) Detail of the $J_0,T$ phase diagram around the
tricritical point.  Full lines are thermodynamic transitions. Also the
transition between the SG (1RSB) and the approximated RS solution for
the FM phase is displayed (double-dotted line) showing no appreciable
difference with the exact one. The dashed line represents the dynamic
PM/SG transition. The dotted bold line represents the FM spinodal
lines both inside the PM and the SG phases. The spinodal of the RS FM
phase is shown as well (smaller dots). }
 \label{fig:PhDi_T_J0_det}
\end{figure}

{\em Paramagnetic phase ---} For low ${\cal P}$ the only phase present
 is completely disordered: all order parameters are zero and we have a
 ``paramagnet'' (PM); for the random laser case this phase
 is expected to correspond to a
 noisy continuous wave emission, and all the mode-phases are
 uncorrelated. Actually, this phase exists for any degree of disorder
 and pumping, yet it becomes thermodynamically sub-dominant as ${\cal
 P}$ (or $\beta$) increases and, depending on the degree of
 disorder, the spin-glass or the ferromagnetic phases take over.
\\ \indent {\em Glassy phase ---} For large disorder, as ${\cal
P}$/$\beta$ grows, a discontinuous transition occurs from the PM to a
spin-glass (SG) phase in which the phases $\phi$ are frozen but do not
display any ordered pattern in space.  First, along the line ${\cal
P}_d=\sqrt{\bar\beta_d/R_J}$, in Fig. \ref{fig:PhDi_P_R}, or at
$T/\sigma_J=1/\bar\beta_d=0.15447$ in Figs. \ref{fig:PhDi_T_J0},
\ref{fig:PhDi_T_J0_det} (dashed lines) a dynamic transition occurs.
Indeed, the lifetime of metastable states is infinite in the
mean-field model and the dynamics gets stuck in the highest lying
excited states.  The thermodynamic state is, however, still PM.
Fig. \ref{fig:PhDi_T_J0_det} displays a detail of the tricritical
region where. There, besides thermodynamic transition lines, we also
plot as dotted curves the lines at which the ferromagnetic phase first
appears as metastable, i.e., the spinodal lines.

\begin{figure}[t!]
\includegraphics[height=.99\columnwidth,angle=270]{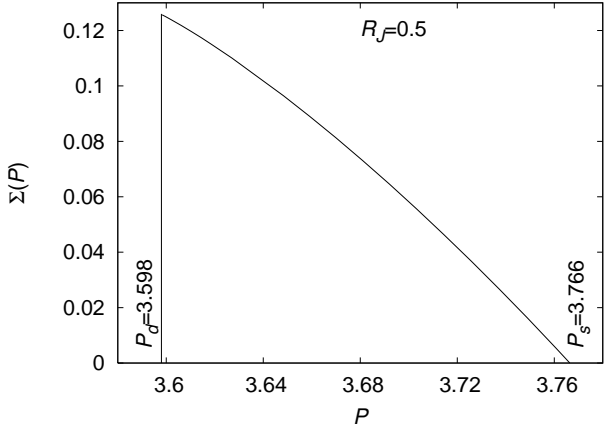}
\caption{Complexity $\Sigma({\cal P})$ of the lowest lying glassy
states in free energy between the values of the pumping rate
corresponding to the dynamic and static transition from the PM to the
SG phase at $R_J=0.5$. }
\label{fig:Sigma_P_SG}
\end{figure}
\begin{figure}[t!]
\includegraphics[height=.99\columnwidth,angle=270]{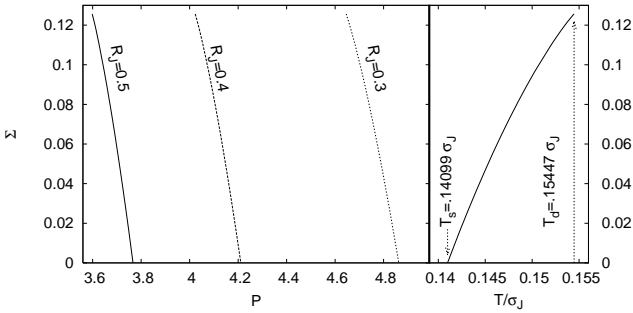}
\caption{Left: Complexity of the lowest lying glassy states in free
energy $\Sigma({\cal P})$ between dynamic and static transition from
the PM to the SG phase along $R_J=0.3,0.4$ and $0.5$ lines. The
qualitative behavior is identical for any $R_J\gtrsim 0.3$. Right:
$\Sigma$ vs. the effective temperature $T$ in $\sigma_J$ units.}
\label{fig:Sigma_T_SG}
\end{figure}
\begin{figure}[t!]
\includegraphics[width=.99\columnwidth]{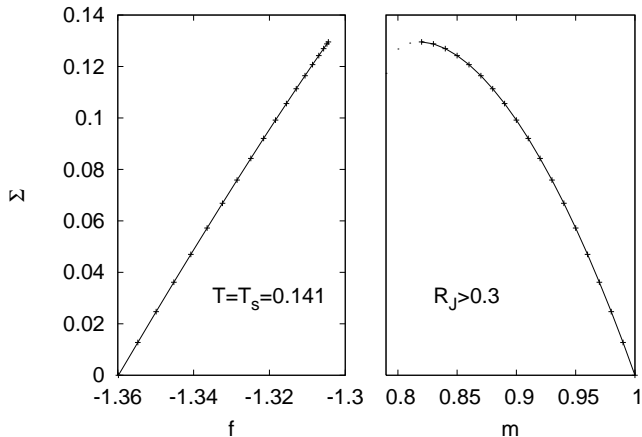}
\caption{$\Sigma(f)$ (left) and $\Sigma(m)$ (right) in the glassy
phase at the static transition effective temperature,
$T=0.14099$. This is the picture holding for any $R_J\gtrsim 0.3$.  }
\label{fig:Sigma_SG_f_Ts}
\end{figure}
\begin{figure}[t!]
\includegraphics[width=.99\columnwidth]{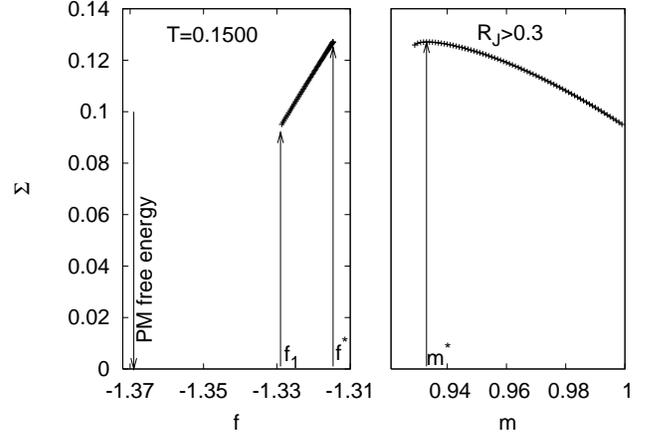}
\caption{$\Sigma(f)$ (left) and $\Sigma(m)$ (right) in the glassy
phase at the static transition effective temperature, $T=0.15<T_d$. The
lowest state free energy of metastable glassy states is denoted by
$f_1$ (i.e., corresponding to $m=1$ in the right panel, see text). The
free energy of maximum complexity is denoted by $f^*$, correspondingly
$m^*$ in the right hand side plot. }
\label{fig:Sigma_SG_f_Ti}
\end{figure}
 In Fig. \ref{fig:Sigma_P_SG} we plot the complexity of the metastable
glassy states of lowest free energy between the dynamic and the static
transition. In the left panel of Fig. \ref{fig:Sigma_T_SG}
$\Sigma({\cal P})$ is displayed for three different values of $R_J$;
the threshold pumping for non-zero minimal complexity grows as the
degree of disorder $R_J$ decreases, as well as the corresponding
${\cal P}$ range. In the right panel $\Sigma(T)$ is plotted and it is 
independent of $R_J$.   In Figs. \ref{fig:Sigma_SG_f_Ts} and
\ref{fig:Sigma_SG_f_Ti} we display two instances of the whole
complexity curve both vs. $f$ and $m$ at $T=T_s$ and at a higher
temperature $T<T_d$.
\\ \indent Across the full line ${\cal P}_s(R_J)=\sqrt{\bar
\beta_s/R_J}$, in Fig. \ref{fig:PhDi_P_R} or, alternatively, across
$T/\sigma_J=1/\bar\beta_s =0.14099$ in Fig.  \ref{fig:PhDi_T_J0}, a
true thermodynamic phase transition from the continuous wave
(paramagnetic) phase to the ``glassy coherent light'' (spin-glass)
phase occurs.  The order parameter $q_1$ (the Edwards-Anderson
parameter $q_{\rm EA}$ \cite{Edwards75}), discontinuously jumps at the
transition from zero $q_{1}>q_0=0$, while $\tilde m=r_0=r_1=r_d=0$
(see Fig. \ref{fig:op}, bottom panel).  The SG phase exists for any value
of $R_J$ and $\bar \beta> \bar \beta_s$.
\\ \indent In the stable SG phase, metastable states (with infinite
lifetime) continue to exist so that the thermodynamic state is
actually unreachable along a standard dynamics starting from random
initial condition.  In Fig. \ref{fig:Sigma_f} we plot the typical
behavior of the complexity versus the single state free energy at
$T/\sigma_J=0.1$, qualitatively identical to the left panel of
Fig. \ref{fig:Sigma_SG_f_Ts} displaying $\Sigma(f)$ at $T=T_s$.

{\em Ferromagnetic phase ---} For weak disorder a random ferromagnetic
 (FM) phase turns out to dominate over both the SG and the PM phases.
 The transition PM/FM line is the standard passive ML threshold (see
 e.g. [\onlinecite{Gordon02,Gordon03}]) and it turns out to be first
 order in the Ehrenfest (i.e., thermodynamic) sense \cite{Leuzzi07,
 Ehrenfest33}. From Fig. \ref{fig:PhDi_P_R} we see that it takes place
 at growing pumping rates ${\cal P}$ for increasing $R_J$ until it
 reaches the tricritical point with the SG phase. In the $(T,J_0)$
 plane it occurs at large - positive - $J_0$, cf.
 Fig.\ref{fig:PhDi_T_J0}
\\ \indent To precisely describe the FM phase in the 1RSB Ansatz we
have to solve eleven coupled integral equations
[Eqs. (\ref{f:speq_q1_R})-(\ref{f:speq_rd}) and Eq. (\ref{f:dphi_dm})
($\Sigma(m;\bm Q_{\rm sp}), \bm\Lambda_{\rm sp})=0$),
cf. Eq. (\ref{f:Sigma})].  
In evaluating their solutions we have to consider that, in the region
where the FM phase is thermodynamically dominant, both the PM and the
SG solutions also satisfy the same set of equations.  Besides,
unfortunately, the basin of attraction of the latter two phases - in
terms of initial conditions - is much broader than the FM one.
Starting the iterative resolution from random initial conditions,
determining the FM transition and spinodal
lines becomes, thus, numerically demanding.
\\ \indent An approximation can be obtained by considering the Replica
Symmetric (RS) solution for the FM phase (FM$_{\rm rs}$). This reduces
the number of independent parameters to seven ($q_1=q_0$,
$r_1^{R,I}=r_0^{R,I}$, $r_d^{R,I}$ and $\tilde m^{R,I}$). The
corresponding transition line is shown as a 
\begin{widetext}
\begin{minipage}{0.99\textwidth}
\centering
\includegraphics[width=\textwidth]{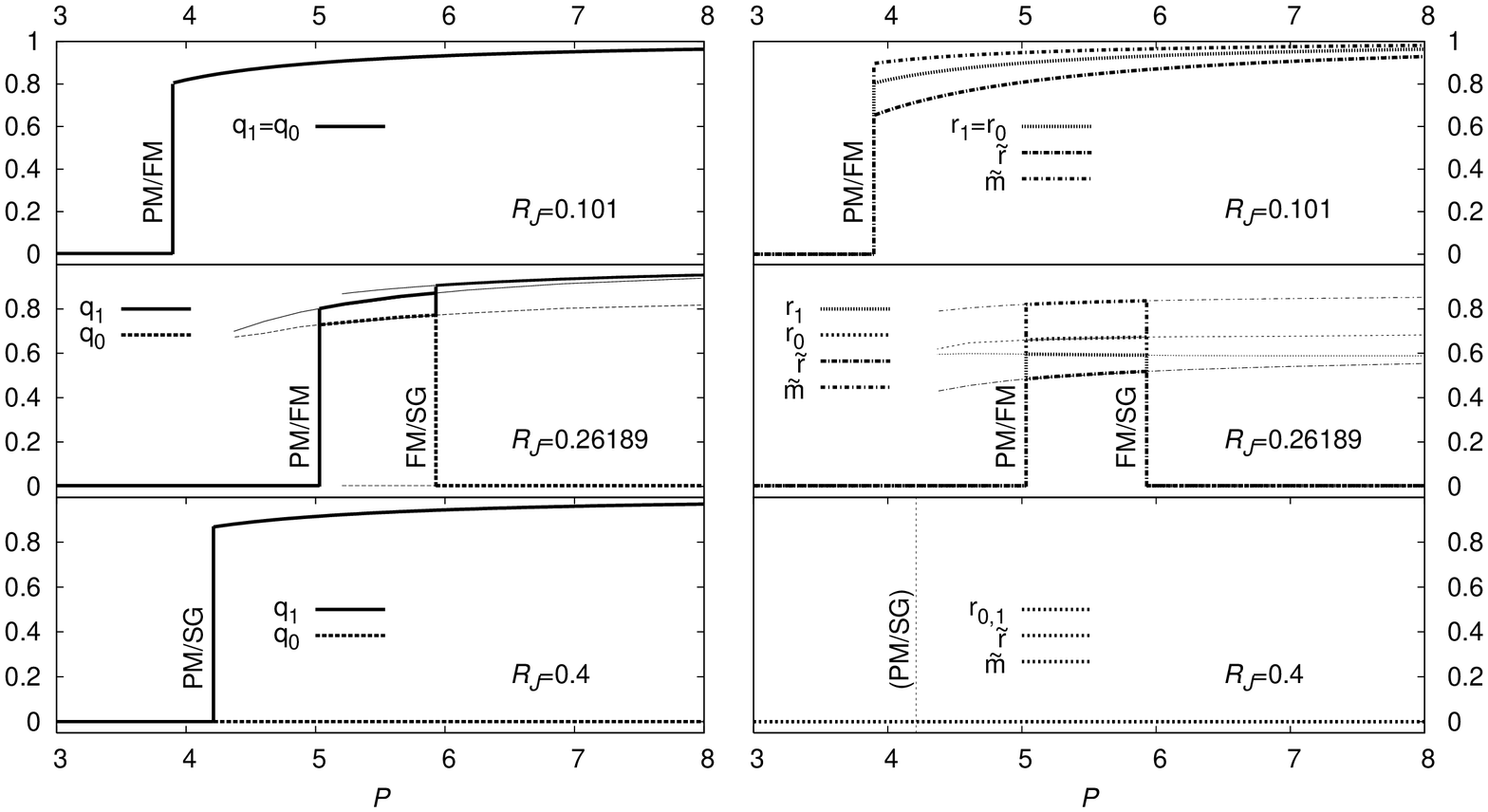}
\figcaption{Discontinuity of the order parameters at the transition
 points for three values of $R_J$. Top Left panel: jump in $q_{0,1}$,
 at the PM/FM transition in
 ${\cal P}$ for small disorder, $R_J\simeq 0.1$; top right:  discontinuities in $r_{0,1}$, ${\tilde r}$ and ${\tilde m}$ at the same transition.   For such small $R_J$
 the replica symmetry breaking is practically invisible: $q_1\simeq
 q_0$, $r_1\simeq r_0$ [to the precision of our computation, ${\cal
 O}(10^{-5})$].  Mid left panel (across tricritical region in
 Fig. \ref{fig:PhDi_P_R}: $q_{0,1}$ vs. ${\cal P}$ at $R_J\simeq 0.26$ where, increasing the pumping
 rate, first a PM/FM transition occurs followed by a FM/SG one. 
 Mid right panel:  $r_{0,1}$, $\tilde r$ and $\tilde
 m$ vs. ${\cal P}$ for the same interval. First
 order transition point are signaled by vertical lines.  Left bottom panel:
 $q_{0,1}$ vs. ${\cal P}$ for large disorder, $R_J=0.4$ across the
 PM/SG random first order transition. Right bottom:  $r_{0,1}$, $\tilde r$ and $\tilde
 m$ are always zero in the SG and in the PM phase.}
\label{fig:op}
\end{minipage}
\end{widetext}
\noindent dashed-dotted line in
Fig. \ref{fig:PhDi_T_J0_det}, where, {\em around the transition}, we
observe no practical difference with the exact SG/FM, even though the
replica symmetry is clearly broken.

In Fig. \ref{fig:op} we show the discontinuous behavior of
the order parameters across various transitions. As disorder is small
(top panel) one can observe that the RSB of the solution representing
the passive mode-locking phase vanishes, at least for what concerns
the limit of precision of our computation. As the degree of disorder
takes values around the tricritical point the RSB is clearly visible
(mid panel), both in the FM and in the SG phases. For increasing
disorder the FM is absent ($R_J\gtrsim 0.263$) and at high pumping/low
temperature only the glassy random laser phase remains.
\begin{figure}
\includegraphics[width=.49\columnwidth]{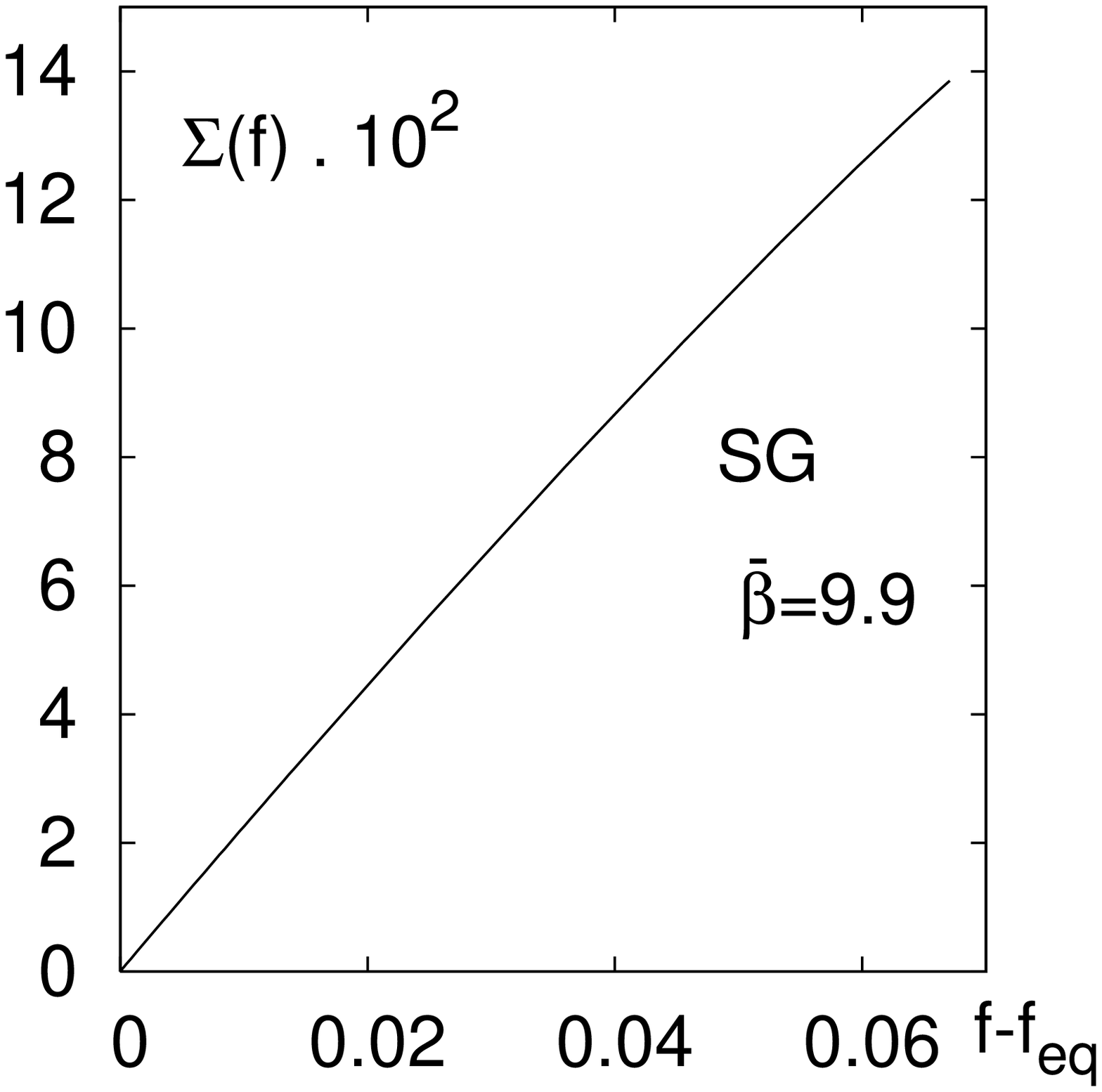}
\includegraphics[width=.49\columnwidth]{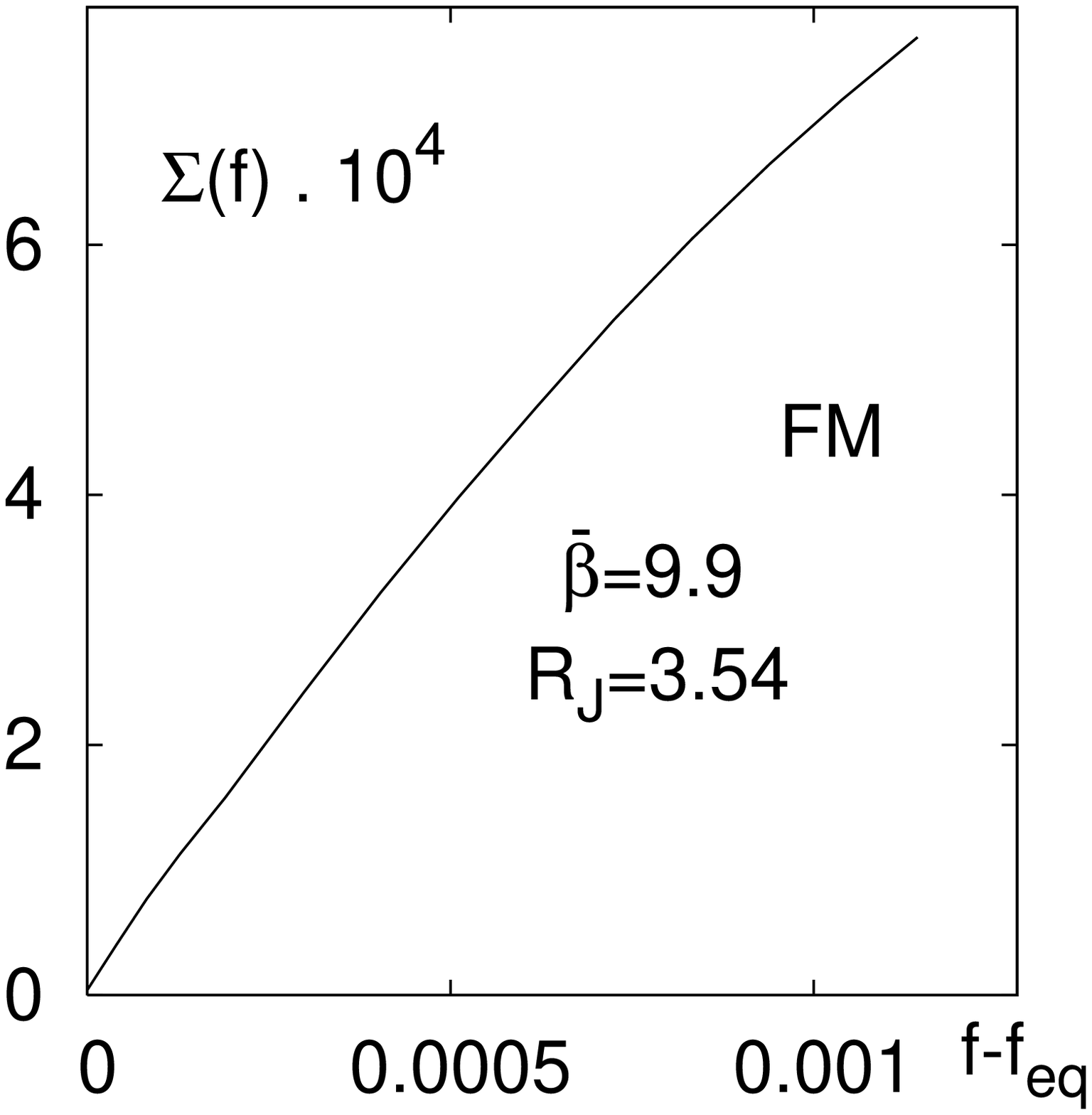}
 \protect\caption{Complexity vs. free energy curve is plotted in the
 SG phase (left) at $T/\sigma_J=0.1$ and in the FM phase (right).}
\label{fig:Sigma_f}
\end{figure}

 We must necessarily implement the 1RSB Ansatz, though, to determine
the not-vanishing extensive complexity which signals the presence of a
large quantity of excited states with respect to ground states and
study its behavior in $T$ and $R_J$. This, as anticipated, also
implies the occurrence of a dynamic transition besides the
thermodynamic one.  In the phase diagrams, Figs. \ref{fig:PhDi_P_R},
\ref{fig:PhDi_T_J0}, \ref{fig:PhDi_T_J0_det}, this takes place between
PM and SG, where the state structure always displays a non-trivial
$\Sigma$, for any $\bar \beta>\bar \beta_d$. Whether an exclusively
dynamic transition can occur as a precursor to the FM phase, as well,
could not be directly established in the present work. Indeed, the
region of expected dynamic transition lies beyond the spinodal FM
line, already very difficult to obtain numerically because of the
competition with the SG and PM solutions.  However, the existence of a
metastable FM phase (cf. spinodal line in
Fig. \ref{fig:PhDi_T_J0_det}) with an extensive complexity, cf. e.g.,
Figs. \ref{fig:Sigma_f} and \ref{fig:Sigma_FM_f}, might well
correspond to an arrest of the dynamic relaxation towards equilibrium
of the system.
\begin{figure}
\includegraphics[width=.99\columnwidth]{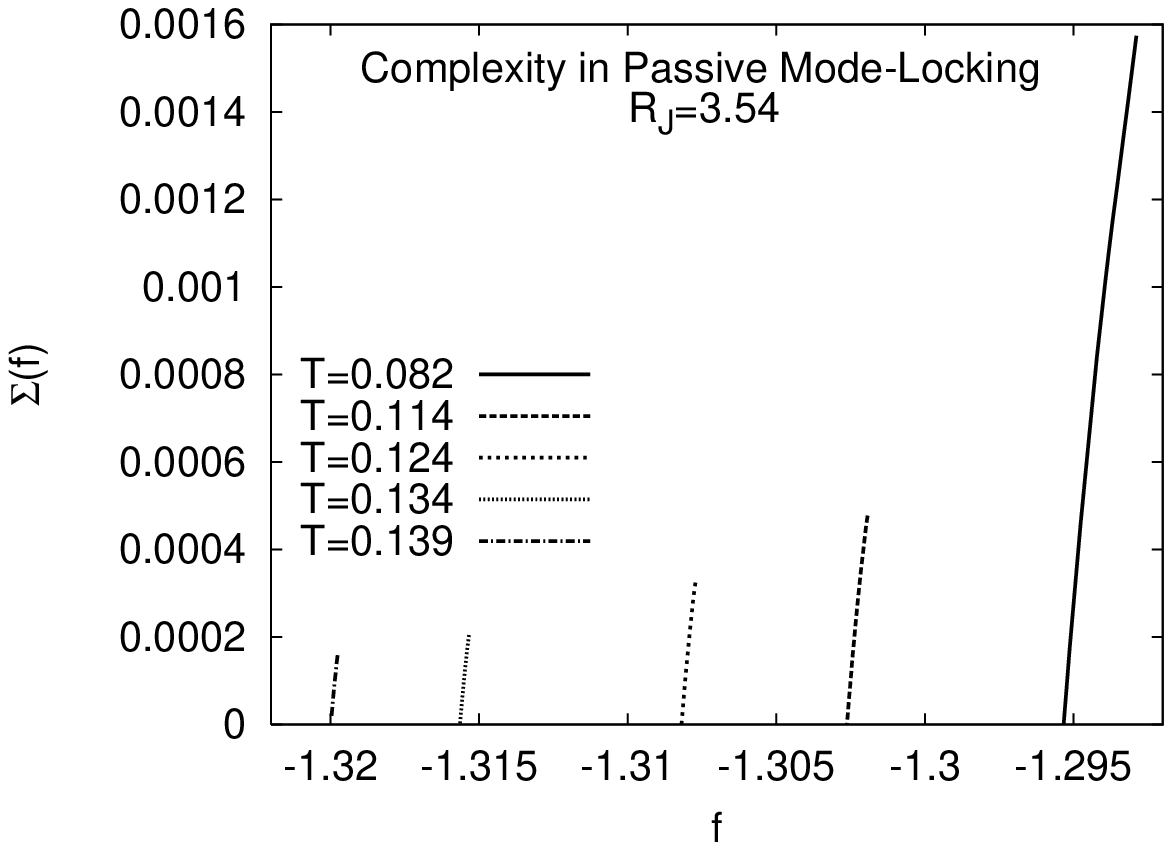}
\caption{Complexity curves of the FM phase at $R_J=3.54$ at
temperatures between $T=0.082\sigma_J$ (right most) and
$T=0.139\sigma_J$ (left most). Both the magnitude of the maximal
complexity and the free energy interval in which $\Sigma(f)>0$
decrease. Notice that the equilibrium free energy decreases as
temperature increases.}
\label{fig:Sigma_FM_f}
\end{figure}
\begin{figure}[t!]
\includegraphics[width=.99\columnwidth]{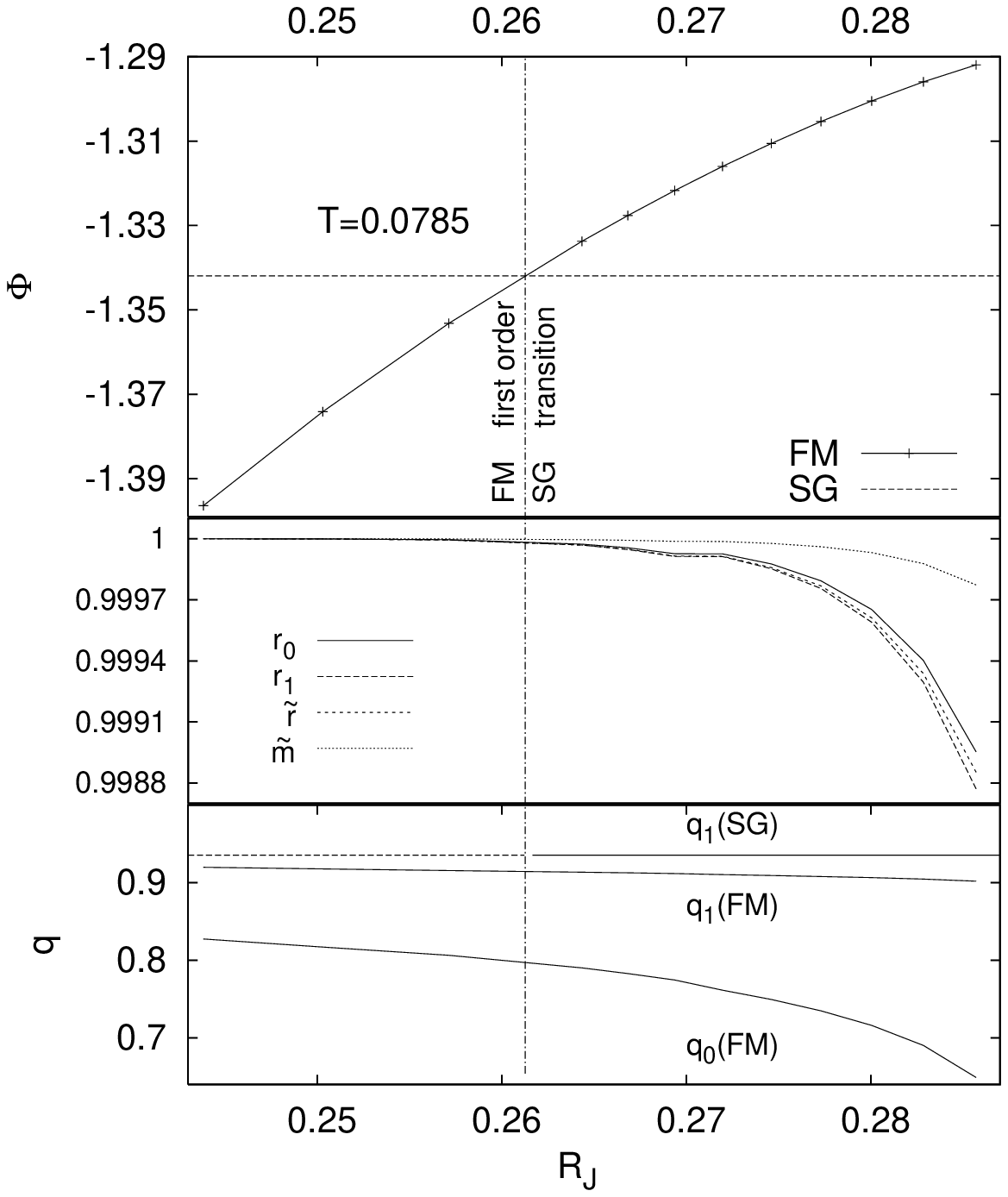}
\caption{Top panel: free energy of the FM and SG phases vs. $R_J$ at
$T=0.0785$.  As the degree of disorder increases the system undergoes
a first order phase transition from a ferromagnetic phase to a
spin-glass. Mid panel: order parameters $r_1,r_0$, $\tilde r$ and the
magnetization $\tilde m$ are shown vs. $R_J$.  Beyond the transition
point their values drop to zero in the SG phase.  Bottom panel: $q$
order parameters for the FM and the SG phase.}
\label{fig:opSGFM}
\end{figure}
\\
\indent
  In the right
inset of Fig. \ref{fig:Sigma_f} we show, e.g., $\Sigma(f)$ in the
FM phase at $(R_J,{\cal P})=(0.28,5.92)$.  This has to be
compared with the SG complexity at the same temperature (left inset
of Fig. \ref{fig:Sigma_f}) that is sensitively larger and does not depend
on the $R_J$: the maximum complexity drops of about
two orders of magnitude at the SG/FM transition, thus
unveiling a corresponding {\em high to low complexity transition}.
\\ \indent In Fig. \ref{fig:opSGFM}, at a relatively low temperature
 $T=0.0785$ we show the behavior of the 1RSB (equilibrium) free energy
 and order parameters across this SG (high complexity)/FM (low
 complexity) transition.  The transition is first order in $R_J$.

\section{Conclusion} 
\label{sec:conclusion}
We have reported on an extensive theoretical treatment of the
thermodynamic and dynamic phases of nonlinear waves in a random
systems. The approach allows to treat nonlinearity and an arbitrary
degree of disorder on the same ground, and predict the existence of
complex coherent phases detailed in a specific phase-diagram.  The
whole theoretical treatment is limited to the quenched-amplitude
approximation, which allows to catch the basic phenomenology and to
demonstrate the existence of phases with a not-vanishing complexity in
a variety of physical systems, and specifically random lasers, finite
temperature BEC and nonlinear optics.  This approximation will be
removed in future works, and novel exotic phases of light in
nonlinear random system will be detailed.

Our theoretical work shows that the interplay of nonlinearity and
disorder leads to the prediction of substantially innovative physical
effects, which bridge the gap between fundamental mathematical models
of statistical mechanics and nonlinear waves.  This allows to identify
frustration and complexity as the leading mechanisms for a coherent
wave regime in nonlinear disordered systems.  Natural extension of
this work will be considering the quantum counterpart of the predicted
transitions, and the analysis of out of equilibrium nonlinear waves
dynamics.

 \acknowledgments The research leading to these results has received
 funding from the European Research Council under the European
 Community's Seventh Framework Program (FP7/2007-2013)/ERC grant
 agreement n. 201766 and from the Italian Ministry of Education,
 University and Research under the Basic Research
 Investigation Fund (FIRB/2008)  program/CINECA grant code RBFR08M3P4.

\appendix
\section{Replica computation of the thermodynamic properties}
\label{app}
The replicated partition function of the system described by the
Hamiltonian ${\cal H}[\{\phi_j\}]$, cf. Eq. (\ref{f:Haphi2}), reads
\BEQ
Z^n_J= \int \prod_{a=1}^n\prod_{j=1}^N d\phi^a_j \exp\left[
-\beta \sum_{a=1}^n{\cal H}[\{\phi_j^a\}]
\right]
\label{f:Z_rep_J}
\EEQ
In order to compute the free energy of the system using the replica trick, cf.
Eq. (\ref{f:Phi_rep}), Eq. (\ref{f:Z_rep_J}) has to be averaged over the
probability distribution of i.i.d. random bonds:
\BEQ
P(J)\equiv \sqrt{\frac{N^3}{2\pi\sigma_J^2}}
\exp\left[
-N^3
\frac{(J-J_0/N^3)^2}
{2 \sigma_J^2}
\right]
\label{f:P_J}
\EEQ

Eq. (\ref{f:Zintegral}), then reads
\BEA 
&&{\overline {Z_J^n}}=\int {\cal D}\phi
\exp\Biggl\{
N\Biggl[
\frac{\beta^2\sigma_J^2}{32}
\sum_{a=1}^n\left(1+\left|\tilde R_{a}(\{\phi\})\right|^4\right) 
\nn
\\
\nn
&&
\hspace*{1.8cm}
+\frac{\beta^2\sigma_J^2}{16}\sum_{a<b}
\left(
\left|Q_{ab}(\{\phi\})\right|^4+\left|R_{ab}(\{\phi\})\right|^4
\right)
\\
&&
\hspace*{4.2cm}
+\frac{\beta J_0}{8}\sum_{a=1}^n \left|M_a(\{\phi\})\right|^4 \Biggr] \Biggr\}
\nn
\\
\EEA
with
\BEA
&&{\cal D}\phi\equiv\prod_{a=1}^n\prod_{j=1}^N d\phi^a_j
\EEA
and
\BEA
&&Q_{ab}(\{\phi\})\equiv \frac{1}{N}\sum_{j=1}^N e^{\imath(\phi_j^a-\phi_j^b)}
\nn
;
\quad
M_{a}(\{\phi\})\equiv \frac{1}{N}\sum_{j=1}^N e^{\imath\phi_j^a}
\\
\label{f:Qphi}
\\
&& R_{ab}(\{\phi\})\equiv \frac{1}{N}\sum_{j=1}^N e^{\imath(\phi_j^a+\phi_j^b)}
;\quad
\tilde R_{a}(\{\phi\})\equiv \frac{1}{N}\sum_{j=1}^N e^{2\imath\phi_j^a}
\nn
\\
\label{f:Mphi}
\EEA
where we used the Euler's formula to represent the cosine and
introduced abbreviations for the quantities
Eqs. (\ref{f:Qphi})-(\ref{f:Mphi}).  
We notice that the matrix $Q_{ab}$ is Hermitian.
A further step is to introduce
extra parameters - that will eventually result as the order parameters
identifying the various phases of the system - by means of the
following identities:
\BEA 
&&1=\prod_{a<b}^{1,n} N \int dq_{ab}
\delta\left[N\left(Q_{ab}(\{\phi\})-q_{ab} \right)\right]
\label{f:intro_qab}
 \\
&&1=\prod_{a<b}^{1,n} N \int dr_{ab}
\delta\left[N\left(R_{ab}(\{\phi\}) -r_{ab}\right)\right]
 \\
&&1=\prod_{a=1}^{n} N \int d\tilde r_{a} \delta\left[N\left(
\tilde R_{a}(\{\phi\})-\tilde r_{a} \right)\right] 
\\
 &&1=\prod_{a=1}^{n} N
\int d\tilde m_{a} \delta\left[N\left(M_{a}(\{\phi\})-\tilde m_{a}
\right)\right] 
\end{eqnarray}
\begin{eqnarray}
&&\delta\left[N\left(Q_{ab}(\{\phi\})-q_{ab}
\right)\right]=\int \frac{d\lambda_{ab}}{2\pi}e^{\Re \left[\bar
\lambda_{ab}N\left(Q_{ab}(\{\phi\})-q_{ab} \right) \right]}
\nn
\\
\\
\nn &&\delta\left[N\left(R_{ab}(\{\phi\})-r_{ab} \right)\right]=\int
\frac{d\mu_{ab}}{2\pi}e^{\Re \left[\bar
\mu_{ab}N\left(R_{ab}(\{\phi\}) -r_{ab}\right) \right]} 
\\ 
\\ \nn
&&\delta\left[N\left(\tilde R_{a}(\{\phi\})-\tilde r_{a}
\right)\right]=\int \frac{d\tilde\mu_{a}}{2\pi}e^{\Re \left[{\bar
{\tilde \mu}}_{a}N\left(\tilde R_{a}(\{\phi\})-\tilde r_{a} \right)
\right]} 
\\ 
\\ \nn &&\delta\left[N\left(M_{a}(\{\phi\})-\tilde m_{a}
\right)\right]=\int \frac{d\nu_{a}}{2\pi}e^{\Re \left[\bar
\nu_{ab}N\left(M_{a}(\{\phi\})-\tilde m_{a} \right) \right]} 
\EEA
The two-index auxiliary variables $q_{ab}, r_{ab}$, defined for
distinct couples of replicas $a$ and $b$, with $a<b$, will be considered in the
following as elements of symmetric matrices [cf. Eqs. (\ref{f:Ansatz_RS})
and (\ref{f:Ansatz_1RSB}) ], i.e., $q_{ab}=q_{ba}$ and $r_{ab}=r_{ba}$. In
particular, since $Q_{ab}(\phi)=\bar Q_{ba}(\phi)$,
Eq. (\ref{f:intro_qab}) implies that $q_{ab}=\bar q_{ab}$ are real
valued.
\\
\indent
Denoting for shortness the sets of
parameters by the ``vectors'' $\mathbf{Q}=\{q,r,\tilde r,\tilde m\}$ and
$\mathbf{\Lambda}=\{\lambda,\mu,\tilde\mu,\nu\}$, this leads to
\BEA 
&&{\overline {Z_J^n}}=\int{\cal
D}\mathbf{Q}~{\cal D}\mathbf{\Lambda}~ e^{-N~n~G\left[
\mathbf{Q},\mathbf{\Lambda}\right]}
\\ \nn &&n~G\left[
\mathbf{Q},\mathbf{\Lambda}\right]=n~A\left[
\mathbf{Q},\mathbf{\Lambda}\right] +\log Z_\phi\left[\mathbf{\Lambda}
\right] 
\\ && n~A \left[\mathbf{Q},\mathbf{\Lambda}\right] \equiv
-\frac{\beta^2\sigma_J^2}{32} \sum_{a=1}^n\Bigl(1+\left| \tilde
r_a\right|^4\Bigr) -\frac{\beta J_0}{8}\sum_{a=1}^n\left| \tilde
m_a\right|^4 \nn 
\\ \nn
&&\hspace*{.5cm}-\frac{\beta^2\sigma_J^2}{16}\sum_{a<b}^{1,n}\Bigl(\left|
q_{ab}\right|^4+\left| r_{ab}\right|^4\Bigr)
-\sum_{a<b}^{1,n}\Re\bigl[
q_{ab}\bar\lambda_{ab}+r_{ab}\bar\mu_{ab}\bigr] \\ &&\hspace*{1cm}-
\sum_{a=1}^n\Re\bigl[\tilde r_a{\bar{\tilde \mu}}_a+\tilde m_a\nu_a
\bigr] 
\EEA
\BEA
&&Z_\phi\left[\mathbf{\Lambda} \right]\equiv
\int\prod_{a=1}^n d\phi_a e^{-\beta{\cal H}_{\rm
eff}[\{\phi\};\mathbf{\Lambda}]}
\label{f:Zphi}
\\
&&-\beta{\cal H}_{\rm eff}[\{\phi\};\mathbf{\Lambda}]\equiv 
\sum_{a<b}^{1,n}\Re\bigl[e^{\imath(\phi_a-\phi_b)}\bar\lambda_{ab}
+e^{\imath(\phi_a+\phi_b)}\bar\mu_{ab}\bigr]
\nn
\\
\label{f:Hphi}
&&\hspace*{2cm}
+\sum_{a=1}^n\Re\bigl[
e^{2\imath \phi_a}{\bar{\tilde \mu}}_a+e^{\imath \phi_a}\bar\nu_a
\bigr]
\EEA
with
\BEA
\nn
&&{\cal D}\mathbf{Q}\equiv \prod_{a<b}^{1,n}N^2 dq_{ab} dr_{ab}\times
\prod_{a=1}^n N^2 d\tilde r_a d\tilde m_a 
\\
\nn
&&{\cal D}\mathbf{\Lambda}\equiv \prod_{a<b}^{1,n} \frac{d\lambda_{ab}}{2\pi}
 \frac{d\mu_{ab}}{2\pi}\times
\prod_{a=1}^n \frac{d\tilde \mu_a}{2\pi}\frac{d\nu_a}{2\pi}
\EEA
The average replicated partition function integral can be estimated by
the saddle point method for large $N$, i.e., by approximating
\BEA
\nn
&&\int {\cal D}[\mathbf{X}]~e^{N~ F[\mathbf{X}]}
\simeq e^{N~F[\mathbf{X}_{\rm sp}]}
\\
\nn
&&\frac{\p F}{\p X_j} 
\bigr|_{\mathbf{X}
=\mathbf{X}_{\rm sp}}=0 \quad \forall~ j=1,\ldots, \mbox{\# parameters}
\EEA
Denoting by $\langle\ldots\rangle_{\rm eff}$ the average over the measure $e^{-\beta{\cal H}_{\rm eff}}$,
the saddle point equations are:
\BEA
\label{f:SPeq_q}
q_{ab}&=&\langle e^{\imath(\phi_a-\phi_b)}\rangle_{\rm eff};\quad
\lambda_{ab} = \frac{\beta^2\sigma_J^2}{4}\left|q_{ab}\right|^2q_{ab}\ \ \ 
\\
r_{ab}&=&\langle e^{\imath(\phi_a+\phi_b)}\rangle_{\rm eff};\quad
\mu_{ab}=\frac{\beta^2\sigma_J^2}{4}\left|r_{ab}\right|^2 r_{ab}\ \ \ 
\\
\nn
&&\forall~a,b=1,\ldots,n;~a < b
\\
\label{f:tildera}
\tilde r_{a\  }&=&\langle e^{2\imath \phi_a}\rangle_{\rm eff};\qquad \quad
\tilde \mu_a=
\frac{\beta^2\sigma_J^2}{8}\left|\tilde r_{a}\right|^2 \tilde r_{a}
\\
\label{f:tildema}
\tilde m_{a}&=&\langle e^{\imath \phi_a}\rangle_{\rm eff};\qquad  \quad \
\nu_a=\frac{\beta J_0}{2}\left|\tilde m_{a}\right|^2 \tilde m_{a}
\\
\nn
&&\forall ~ a=1,\ldots,n
\EEA
The diagonal values of the overlap matrices are set to zero.
Eventually, according to Eq. (\ref{f:Phi_rep}), one has
\BEQ \beta\Phi=-\frac{1}{N}\lim_{n\to
0}\frac{e^{-nNG\left[\mathbf{Q}_{\rm sp}, \mathbf{\Lambda}_{\rm
sp}\right]}-1}{n}=\lim_{n\to 0}G\left[\mathbf{Q}_{\rm sp},
\mathbf{\Lambda}_{\rm sp}\right] \EEQ
The parameters with a single replica index turn out not to depend on
the specific replica.  Indeed,
Eqs. (\ref{f:tildera})-(\ref{f:tildema}) might in principle be
obtained by perturbing the original Hamiltonian with a small field
coupled to a local function of the planar, XY, spins $S=e^{\imath
\phi}$, independently from the possible introduction of replicas. 

If
the perturbation is $-k\sum_{j=1}^N S_j^2$ we obtain
\BEA \tilde r_d=-\frac{1}{N}\frac{\p {\overline{\log Z_J}}}{\p k}
=\frac{1}{N}\sum_{j=1}^N{\overline{ S_j^2 }}\, \to \, {\overline{\langle
S\rangle}} \quad \mbox{(as $N\to\infty$)}
\nn
\\
\label{f:rd_RS}
\EEA
that is valid {\em for any} replica, and, therefore independent from
any replica index: $\tilde r_d = \tilde r_a$, $\forall a=1,\ldots, n$.
In the replica formalism, the same quantity can equivalently be
written, as 
\BEQ \tilde r_d=-\lim_{n\to 0}\frac{1}{nN}\frac{\p
{\overline {Z_J^n}}}{\p k} = \lim_{n\to
0}\frac{1}{n}\sum_{a=1}^n\tilde r_a
%\frac{1}{N}\sum_{j=1}^N{\overline{ \left(S_j^a\right)^2  }}=
\nn
\EEQ
 and this trivially leads to the identification 
\BEQ
\tilde r_a = \lim_{n\to 0}\frac{1}{n}\sum_{a=1}^n\tilde r_a  = \tilde r_d
\EEQ
Similarly, perturbing Eq. (\ref{f:Haphi2}) with
 $-h\sum_{j=1}^N S_j$ we get
\BEA
\tilde m&=&-\frac{1}{N}
\frac{\p  {\overline {\log Z_J}}}
{\p h}
 =\frac{1}{N}\sum_{j=1}^N{\overline{ S_j  }}
\label{f:mag_RS}
\\
\nn
\tilde m&=&-\lim_{n\to 0}\frac{1}{nN}
\frac{\p  {\overline {Z_J^n}}}
{\p h} 
= \lim_{n\to 0}\frac{1}{n}\sum_{a=1}^n\tilde m_a
%\frac{1}{N}\sum_{j=1}^N{\overline{ S_j^a  }}
\EEA

Though no external ad hoc perturbation can be applied to the
Hamiltonian Eq. (\ref{f:Hphi}) to reproduce two indices quantities, the same
symmetry should apply, since all replicas of the original problem
were introduced in the same way: the
system is symmetric under replica exchange.  This is called the
replica symmetric (RS) Ansatz.
\BEQ
q_{ab}=q ~~~\forall a\neq b~;\qquad r_{ab}=r~~~\forall a\neq b
\label{f:Ansatz_RS}
\EEQ

\subsection{Replica Symmetric Ansatz}
In this Ansatz Eqs. (\ref{f:Zphi},\ref{f:Hphi}) 
become

\BEA
&&Z_\phi^{\rm RS}=\int\prod_{a=1}^nd\phi_a e^{-\beta{\cal H}_{\rm eff}[\{\phi_a\}]}
\\
&&\beta{\cal H}_{\rm eff}=
\frac{\lambda^R}{2}\left(n-\left|\sum_{a=1}^ne^{\imath \phi_a}\right|^2\right)
\\
\nn
&&-\Re\left[
\frac{\bar\mu}{2}\left(\sum_{a=1}^ne^{\imath\phi_a}\right)^2+
\left(\bar{\tilde{\mu}}-\frac{\bar\mu}{2}
\right)\sum_{a=1}^n e^{2\imath\phi}+\bar\nu\sum_{a=1}^ne^{\imath\phi}
\right]
\EEA
The second term in the rhs can be rewritten as
\BEA
&&\hspace*{-1cm}\Re\left[
\frac{\bar\mu}{2}\left(\sum_{a=1}^ne^{\imath\phi_a}\right)^2\right]=
\Re\left[
\frac{\mu}{2}\left(\sum_{a=1}^ne^{-\imath\phi_a}\right)^2\right]
\\
\nn
&&=
\Re\left[\frac{1}{4}\left(
\sqrt{\bar\mu}\sum_{a=1}^ne^{\imath\phi_a}+
\sqrt{\mu}\sum_{a=1}^ne^{-\imath\phi_a}
\right)^2\right]
\\
\nn
&&\ \ \ -\frac{|\mu|}{2}\left| \sum_{a=1}^ne^{\imath\phi_a}\right|^2
\EEA
The squared terms in the exponent of the integrand can be linearized by using
\BEA
e^{|w|^2/2} = \int\frac{d\zeta^Rd\zeta^I}{2\pi}e^{-|\zeta|^2/2}
e^{\Re(\bar\zeta w)}
\\
e^{w_R^2/2}= \int\frac{dx}{\sqrt{2\pi}}e^{-x^2/2}e^{x w_R}
\EEA
\noindent thus yielding
\BEA
&&\hspace*{-1cm}Z_\phi^{\rm RS}=\int {\cal D}p(x){\cal D}p(\zeta^R){\cal D}p(\zeta^I)
%\times\\
%\nn
%&&\hspace*{3cm}
\biggl[
\int _0^{2\pi} d\phi e^{{\cal L}(\phi;x,\zeta)}
\biggr]^n
\\
&&\hspace*{-1cm}{\cal L}(\phi;x,\zeta)\equiv
\Re\biggl[
e^{\imath\phi}
\left(\bar\zeta\sqrt{\lambda-|\mu|}+
x{\sqrt{2\bar\mu}}+\bar \nu\right)
\label{f:L_RS}
\\
\nn
&&\hspace*{4cm}+
e^{2\imath\phi}
\left(
{\bar{\tilde{\mu}}}-\frac{\bar\mu}{2}\right)\biggr]
\\
&&\hspace*{-1cm}{\cal D}p(w)=\frac{dw}{\sqrt{2\pi}}e^{-w^2/2}
\EEA
The replicated free energy eventually reads:
\BEA
\label{f:Phi_RS}
&&\beta\Phi=-\frac{\beta^2\sigma_J^2}{32}\left[1-|q|^4-|r|^4+|\tilde r|^4\right]
\\
&&\nn-\frac{\beta J_0}{8}|\tilde m|^4+\frac{\lambda^R}{2}(1-q^R)-\frac{1}{2}\Re\left[\bar\mu r-
2{\bar{\tilde{\mu}}}{\tilde r}-2{\bar \nu}\tilde m\right]
\\
\nn
&&-\int {\cal D}p(x){\cal D}p(\zeta^R){\cal D}p(\zeta^I)\log\int _0^{2\pi} d\phi ~e^{{\cal L}(\phi;x,\zeta)}
\EEA

Deriving w.r.t. to $Q$'s parameter we obtain the specification of Eqs. 
(\ref{f:SPeq_q}-\ref{f:tildema}) for the replica overlap parameters $q$, $r$ and
 for $\tilde r$ and $\tilde m$
\BEA
\lambda& =& \frac{\beta^2\sigma_J^2}{4} q^3\ \ \ 
\\
\mu&=&\frac{\beta^2\sigma_J^2}{4}\left|r\right|^2 r\ \ \ 
\\
\tilde \mu&=&\frac{\beta^2\sigma_J^2}{8}\left|\tilde r\right|^2 \tilde r
\nn
\\
\nu&=&\frac{\beta J_0}{2}\left|\tilde m\right|^2 \tilde m
\nn
\EEA

Taking the derivative of $\beta\Phi$ in Eq. (\ref{f:Phi_RS}) 
 w.r.t.  $\tilde\mu$ and $\nu$ we obtain 
\BEA
\label{f:speq_rd_RS}
\tilde r_d&=&\langle\langle e^{2\imath\phi}\rangle_{\cal L}\rangle_{x,\zeta}
\\
\tilde m&=&\langle\langle e^{\imath\phi}\rangle_{\cal L}\rangle_{x,\zeta}
\label{f:speq_mag}
\EEA
where we define
\BEA
\langle\ldots\rangle_{\cal L}
\equiv\frac{\int_0^{2\pi}d\phi \ldots ~e^{{\cal L}(\phi;x,\zeta)}}
{\int_0^{2\pi}d\phi~e^{{\cal L}(\phi;x,\zeta)}}
\EEA

 Deriving $G$ w.r.t. $\lambda$ and $\mu$ and equating to zero
we obtain 
\BEA
q^R&=&\left<c_{\cal L}^2+s_{\cal L}^2\right>_{x,\zeta}
\label{f:speq_q}
\\
\label{f:speq_r}
r &= &\left<c_{\cal L}^2-s_{\cal L}^2
+2\imath~ c_{\cal L} s_{\cal L}\right>_{x,\zeta}
\\
\nn
&&c_{\cal L}\equiv \langle \cos\phi\rangle_{\cal L}
\qquad s_{\cal L}\equiv \langle \sin\phi\rangle_{\cal L}
\EEA
after having integrated by part in the Gaussian measures.  To help the
non-expert reader to easily derive the self-consistency equations we
exemplify the calculation of Eq. (\ref{f:SPeq_q}).
\BEA
&& 2\frac{\p G}{\p \lambda^R} = 0 = 1-q^R
\\
\nn
&&\ \ \ -
%\int {\cal D}p(x)\int {\cal D}p(\zeta^R)\int {\cal D}p(\zeta^I)
\left<\left(\zeta^R c_{\cal L}+\zeta^I s_{\cal L}
\right)\right>_{x,\zeta}
/\sqrt{\lambda^R-|\mu|}
\label{f:dGdla}
\EEA 
The latter term can be simplified by integrating by part 
\BEQ
\int_{-\infty}^\infty {\cal D}p(y)~ y~ F(y) = \int _{-\infty}^\infty
{\cal D}p(y)~\frac{\p F(y)}{ \p y} 
\label{f:bypart}
\EEQ 
with $y=\zeta^R,\zeta^I$ in
Eq. (\ref{f:dGdla}), yielding 
\BEA &&\left< \zeta^R c_{\cal L}+\zeta^I
s_{\cal L} \right>_{x,\zeta}=\sqrt{\lambda^R-|\mu|}\times \\ \nn
&&\quad\left<\cos^2\phi-c_{\cal L}^2 +\sin^2\phi-s_{\cal L}^2
\right>_{x,\zeta} \EEA
The self-consistency equation can thus be rewritten as,
cf. Eq. (\ref{f:speq_q1_R}),
\BEA &&1-q^R=1-\left<c_{\cal L}^2+s_{\cal L}^2\right>_{x,\zeta} \nn \\
\nn && q^R=\left<c_{\cal L}^2\right>_{x,\zeta} +\left<s_{\cal
L}^2\right>_{x,\zeta} \EEA
We recall that since
$q_{ab}$ is real, and so is $\lambda_{ab}$,
cf. Eq. (\ref{f:SPeq_q}), in the RS Ansatz the equations $q^I=\lambda^I=0$.

  Before deriving Eq. (\ref{f:speq_r}), we
rewrite the part of Eq. (\ref{f:L_RS}) involving the integrating
variable $x$ as: 
\BEA && \Re\biggl[ e^{\imath\phi}
x{\sqrt{2\bar\mu}}\biggr]= \\ \nn && \qquad
x\sqrt{|\mu|}\Biggl(\cos\phi\sqrt{1+\frac{\mu^R}{|\mu|}}+\sin\phi
\sqrt{1-\frac{\mu^R}{|\mu|}}\Biggr) 
\label{f:L_RS_x}
\EEA
 In determining the above expression one can use, e.g., the trigonometric
law of tangents to yield 
\BEQ
\frac{1}{2}\arctan \frac{\mu^I}{\mu^R} 
= \arctan \sqrt{\frac{1-\mu^R/|\mu|}{1+\mu^R/|\mu|}}
\EEQ
 and the relationships between
trigonometric and inverse trigonometric functions: 
\BEA
\nn
&&\sin[\arctan(\theta)] = \frac{\theta}{\sqrt{1+\theta^2}}
\\
&&\cos[\arctan(\theta)] = \frac{1}{\sqrt{1+\theta^2}}
\nn
\EEA
Using Eq. (\ref{f:L_RS_x}), together with Eqs. (\ref{f:speq_rd_RS})
and (\ref{f:speq_q}), we have:
\BEA
&& 2\frac{\p G}{\p \mu^R} = 0 = r_d^R-r^R+\frac{\mu^R}{|\mu|}(1-q)
\\
\nn
&&\ \ \ -
%\int {\cal D}p(x)\int {\cal D}p(\zeta^R)\int {\cal D}p(\zeta^I)
\left<x \left(c_{\cal L}\sqrt{1+\frac{\mu^R}{|\mu|}}
+s_{\cal L}\sqrt{1-\frac{\mu^R}{|\mu|}}
\right)\right>_{x,\zeta}
/\sqrt{|\mu|}
\label{f:dGdmur}
\EEA 
Integrating by part with Eq. (\ref{f:bypart}), $y=x$, we find
\BEA
&&\left<x \left(c_{\cal L}\sqrt{1+\frac{\mu^R}{|\mu|}}
+s_{\cal L}\sqrt{1-\frac{\mu^R}{|\mu|}}
\right)\right>_{x,\zeta} 
\\
&&
\nn
\hspace*{2cm } = \sqrt{|\mu|}\Biggl<
\biggl[
\langle\cos 2\phi\rangle_{\cal L}
-c_{\cal L}^2+s_{\cal L}^2
\\
\nn
&&\hspace*{3.9cm }
+\frac{\mu^R}{|\mu|}\left(
1-c_{\cal L}^2-s_{\cal L}^2
\right)
\biggr]
\Biggr>_{x,\zeta}
\EEA
and eventually one obtains the real part of Eq. (\ref{f:speq_r}).
The imaginary part of the self-consistency equation for $r$ is
analogously determined from $\frac{\p G}{\p \mu^I} = 0$.

Above a given critical temperature (depending on $J_0$) the solution
to
Eqs. (\ref{f:speq_rd_RS},\ref{f:speq_mag},\ref{f:speq_q},\ref{f:speq_r})
is paramagnetic, i.e., $ q=r=\tilde m=\tilde r=0$ and the free energy is
\BEQ
\beta\Phi_{PM} = -\frac{\bar\beta^2}{32}-\log 2\pi
\EEQ
Below $T_c(J_0)$,
depending on the value of $J_0$ the solution can either be
ferromagnetic $\tilde m\neq 0$ or spin-glass $\tilde m=0$. The latter
solutions are, however, not stable against fluctuations in the space
of replica overlaps \footnote{The case at $J_0=0$ was explicitly
considered in Ref. [\onlinecite{Angelani06}].} and, thus, we have to
try an Ansatz different from Eq. (\ref{f:Ansatz_RS}) to provide a
self-consistent thermodynamics.
 
\subsection{One step of Replica Symmetry Breaking}

In order to obtain a thermodynamically consistent result the symmetry
cannot be conserved. We are in presence of a spontaneous Replica
Symmetry Breaking. The way to break the symmetry must be a-priori
hypothesized, since there has been found, so far, no way to deduce it.
The correct way to express the elements of the overlap matrices is
called Parisi Ansatz \cite{Parisi79,Parisi80} and, depending on the
kind of system, can consist of one or more RSB's.  According to what
happens in other spin models with $p$-body quenched random
interactions ($p$ being larger than $2$), the right Ansatz for the
matrices of our model is the one-step RSB, that is, we have a $n\times
n$ matrix divided in square blocks of $m\times m$ elements 
\BEA
q_{ab}=q_1;\quad r_{ab}=r_1\ \ \ \mbox{if} ~ I\left(\frac{a}{m}\right)
= I\left(\frac{b}{m}\right) \\ q_{ab}=q_0 ; \quad r_{ab}=r_0\ \ \
\mbox{if} ~I\left(\frac{a}{m}\right) \neq I\left(\frac{b}{m}\right)
\label{f:Ansatz_1RSB}
\EEA 
For instance, for $n=6$ and $m=3$.
\begin{center}
 \begin{math}
 q_{\left(\alpha\beta\right)}=
 \left(
 \begin{array}{clrrrr}
 0 & q_1  & q_1  & q_0 & q_0  & q_0 \\
 q_1 & 0  & q_1  & q_0 & q_0  & q_0 \\
 q_1 & q_1  & 0  & q_0 & q_0  & q_0 \\
 q_0 & q_0  & q_0  & 0 & q_1  & q_1 \\
 q_0 & q_0  & q_0  & q_1 & 0  & q_1 \\
 q_0 & q_0    & q_0  & q_1 & q_1  & 0
 \end{array}
 \right)
 \end{math}
 \end{center}

The one replica index observables are instead still RS, as exemplified in Eqs.
(\ref{f:rd_RS}-\ref{f:mag_RS}).
Now, let us write the "vectorial" replica index
\BEA
a\to \mathbf{a}=(a_1,a_2)
\nn
\\
\phi_{a}\to \phi_{\mathbf{a}}=\phi_{a_1,a_2}
\nn
\\
\nn
\sum_{a=1}^n O_a=\sum_{a_1=1}^{n/m}\sum_{a_2=1}^m O_{a_1a_2}
\EEA
Take a 1RSB matrix $K_{ab}$ and two replicated observables $g_a$ and $h_a$. The following expressions hold for the
sum of a generic product
\BEA
&&2 \sum_{a<b} K_{ab} g_a h_b =  K_{ab}g_ah_b
\\
\nn
&=&K_1\sum_{a_1=1}^{n/m}\sum_{a_2\neq b_2}^{1, m} g_{a_1a_2} h_{a_1b_2}
+K_0\sum_{a_1\neq b_1}^{1,n/m}\sum_{a_2b_2}^{1,m} g_{a_1a_2}
h_{b_1b_2}
\\
\nn
&=&K_1\sum_{a_1=1}^{n/m}\sum_{a_2,b_2}^{1,m} g_{a_1a_2} h_{a_1b_2}
-K_1\sum_{a_1=1}^{n/m}\sum_{a_2=1}^m g_{a_1a_2}h_{a_1a_2}
\\
\nn
&&\hspace*{-.5cm}+K_0\sum_{a_1,b_1}^{1,n/m}\sum_{a_2,b_2}^{1,m} g_{a_1a_2} h_{b_1b_2}
-K_0\sum_{a_1=1}^{n/m}\sum_{a_2,b_2}^{1,m} g_{a_1a_2} h_{a_1b_2}
\EEA
If we take $g=\bar h$ 
\BEA
2 \sum_{a<b} K_{ab} |g_a|^2 &=& (K_1-K_0)\sum_{a_1=1}^{n/m}\left|
\sum_{a_2=1}^{m}g_{a_1a_2}
\right|^2
\nn
\\
\label{f:1rsb_mod}
&&\hspace*{-.5cm}-K_1\sum_{a=1}^n\left|g_{a}\right|^2
+K_0\left|\sum_{a=1}^n g_{a}\right|^2
\EEA
 
 If we take $g= h$ 
\BEA
2 \sum_{a<b} K_{ab} |g_a|^2 &=& (K_1-K_0)\sum_{a_1=1}^{n/m}\left(
\sum_{a_2=1}^{m}g_{a_1a_2}
\right)^2
\nn
\\
\label{f:1rsb_squ}
&&\hspace*{-.5cm}-K_1\sum_{a=1}^n\left(g_{a}\right)^2
+K_0\left(\sum_{a=1}^n g_{a}\right)^2
\EEA
Substituting into Eqs. (\ref{f:Zphi},\ref{f:Hphi}) both Eq. (\ref{f:1rsb_mod}) - with $K=\bar\lambda$ and 
$g=\bar h = e^{\imath\phi}$ -
and  Eq. (\ref{f:1rsb_squ}) - with $K=\bar\mu$ and $g=h=e^{\imath \phi }$ - one obtains
\BEA
Z_{\phi}^{1RSB}&=&\int \prod_{a_1=1}^{n/m}\prod_{a_2=1}^m
e^{-\beta{\cal H}_{\rm eff}[\{\phi_{a_1a_2}\}}] 
\label{f:Z1rsb}
\\
\beta{\cal H}_{\rm eff}& =&\Re \Bigl[
-\frac{\Delta\bar\lambda}{2}\sum_{a_1=1}^{n/m}
\left|\sum_{a_2=1}^m e^{\imath \phi_{a_1a_2}}\right|^2
\\
\nn
&&\qquad-n\frac{\bar\lambda_1}{2}
+\frac{\bar \lambda_0}{2}\left|\sum_{a=1}^ne^{\imath\phi_a}\right|^2
\\
\nn
&&\qquad+\frac{\Delta\bar\mu}{2}\sum_{a_1=1}^{n/m}
\left(\sum_{a_2=1}^m e^{\imath \phi_{a_1a_2}}\right)^2
\\
\nn
&&\qquad-\frac{\bar\mu_1}{2}\sum_{a=1}^ne^{2\imath\phi}
+\frac{\bar \mu_0}{2}\left(\sum_{a=1}^ne^{\imath\phi_a}\right)^2
\\
\nn
&&\qquad\quad+{\bar{\tilde \mu}}\sum_{a=1}^ne^{2\imath\phi}
+\bar\nu\sum_{a=1}^ne^{\imath\phi}
\Bigr]
\\
\nn
&&\Delta\bar\lambda=\bar\lambda_1-\bar\lambda_0
\qquad \Delta\bar\mu=\bar\mu_1-\bar\mu_0
\EEA

Using the identities
\BEA
&&\hspace{-.5cm}\Re[(ab)^2]=\Re[(\bar{a}\bar{b})^2]=\Re\left[\frac{(ab+\bar{a}\bar{b})^2} {2}-|a||b|\right]
\\
&&\hspace{-.5cm}e^{A_R|g|^2/2}=\int\frac{d\zeta_Rd\zeta_I}{2\pi}e^{-|\zeta|^2/2}e^{\Re[\bar\zeta\sqrt{A_R}g]}
\\
&&\hspace{-.5cm}e^{\Re[A~ g^2]/2}=\int_{-\infty}^\infty\frac{dx}{\sqrt{2\pi}}e^{-x^2/2}e^{x\Re[ \sqrt{A} g]}
\EEA
where $A$ is complex and $A_R$ is real,
we can linearize the dependence on $e^{\imath \phi}$ in the partition function Eq. (\ref{f:Z1rsb}) using Gaussian integral expressions:
\BEA
&&\hspace*{-.7cm}
\Re\left[\frac{\bar\lambda_0}{2}\left|\sum_{a=1}^ne^{\imath\phi_a}\right|^2+\frac{\bar\mu_0}{2}\left(\sum_{a=1}^ne^{\imath\phi_a}\right)^2\right]
\\
\nn
&&=
\log\int{\cal D}[\mathbf{0}]
\exp\Re\Bigl[\bar \zeta_0\sqrt{\lambda_0^R-|\mu_0|}\sum_{a=1}^ne^{\imath\phi_a}
%\exp\Re\left[\bar\zeta_0\sqrt{\lambda^R_0}\sum_{a=1}^n e^{\imath\phi_a} \right]
\\
&&+\frac{x_0}{\sqrt{2}}\left(
\sqrt{\bar\mu_0}
\sum_{a=1}^n e^{\imath\phi_a}+
\sqrt{\mu_0}
\sum_{a=1}^n e^{-\imath\phi_a}
\right)
 \Bigr]
 \nn
 \\
&&
%\sum_{a_1=1}^{n/m}
\hspace*{-.7cm}
\Re\left[\frac{\Delta\bar\lambda}{2}
\left| \sum_{a_2=1}^m e^{\imath\phi_{a_1a_2}}\right|^2
+\frac{\Delta\bar\mu}{2}
\left( \sum_{a_2=1}^m e^{\imath\phi_{a_1a_2}}\right)^2
\right]
\\
\nn
&&=\log 
%\prod_{a_1=1}^{n/m}
\int{\cal D}[\mathbf{1}]
\exp\Re\Bigl[
\bar\zeta_1
\sqrt{\Delta\lambda^R-|\Delta\mu|}
\sum_{a_2=1}^m e^{\imath\phi_{a_1a_2}}
\\
&&+
\frac{x_1}{\sqrt{2}}
\Bigl(\sqrt{\Delta\bar\mu}\sum_{a_2=1}^m e^{\imath\phi_{a_1a_2}}
%\nn
%\\
%&&\qquad\qquad\quad\quad
+\sqrt{\Delta\mu}\sum_{a_2=1}^m e^{-\imath\phi_{a_1a_2}}\Bigr)
 \Bigr]
 \nn
\EEA
where we defined the Gaussian measures:
\BEA
&&{\cal D}p(\zeta_k^{R,I}) = \frac{d\zeta_k^{R,I}}{\sqrt{2\pi}}
e^{-\left(\zeta_k^{R,I}\right)^2/2}
\\
&&{\cal D}p(x_k) = \frac{dx_k}{\sqrt{2\pi}}
e^{-x_k^2/2}
\\
&&{\cal D}[\mathbf{k}]={\cal D}p(\zeta_k^R){\cal D}p(\zeta_k^I) {\cal D}p(x_k) 
\EEA

Eq. (\ref{f:Z1rsb}) becomes
\BEA
&&\hspace*{-.3cm}Z_\phi^{1RSB}=e^{-n\lambda_1^R/2}\times
\\
\nn
&&\qquad \int {\cal D}[\mathbf{0}]\prod_{a_1=1}^{n/m}\Bigl\{
\int {\cal D}[\mathbf{1}]
\int{\cal D}\phi
\prod_{a_2=1}^m e^{{\cal L}(\phi_{a_1a_2};\mathbf{0},\mathbf{1})}
\Bigr\}
\\
&&\hspace*{-.3cm}{\cal L}(\psi;\mathbf{0},\mathbf{1})\equiv \Re\Bigl[e^{\imath\psi}\Bigl(\bar\zeta_1
\sqrt{\Delta\lambda^R-|\Delta\mu|}
\\
\nn
&& +\bar\zeta_0
\sqrt{\lambda_0^R-|\mu_0|}+2x_1\sqrt{\frac{\Delta\bar\mu}{2}}
+2x_0\sqrt{\frac{\bar\mu_0}{2}}+\bar \nu\Bigr)
\\
\nn
&&\hspace*{4cm} + e^{2\imath\psi}\left({\bar{\tilde{\mu}}}-\frac{\bar\mu_1}{2}\right)
\Bigr]
\EEA
cf. Eq. (\ref{f:calL}).
In the $n\to 0$ limit the 'phase contribution' to the replicated free energy is:
\BEA
&&-\lim_{n\to 0}\frac{1}{n}\log Z_\phi^{1RSB}=\frac{\lambda_1^R}{2}
\\
\nn
&&\qquad-\frac{1}{m}
\int{\cal D}[\mathbf{0}]\log\int{\cal D}[\mathbf{1}]\left[\int d\phi ~e^{{\cal L}(\phi;\mathbf{0},\mathbf{1})}\right]^m
\EEA
and
the free energy is
\BEA
&&\beta\Phi = \lim_{n\to 0}G_{\rm 1RSB}[\mathbf{Q}_{\rm sp};\mathbf{\Lambda}_{\rm sp}]
\\
\nn
&&\ \ =
-\frac{\beta^2\sigma_J^2}{32}\Bigl[
1-(1-m)\left(|q_1|^4+|r_1|^4\right)
\\
\nn
&&\quad-m\left(
|q_0|^4+|r_0|^4\right)+|\tilde r|^4\Bigr]
\\
\nn
&&\quad-\frac{1}{2}\Re\Bigl[
(1-m)(\bar\lambda_1q_1+\bar\mu_1r_1)+
m(\bar\lambda_0q_0+\bar\mu_0r_0)\Bigr]
\\
\nn
&&\quad -\frac{\beta J_0}{8}|\tilde m|^4
+\Re\Bigl[
{\bar{\tilde\mu}}\tilde r+\bar\nu\tilde m
\Bigr]
-\lim_{n\to 0}\frac{1}{n}\log Z^{\rm 1RSB}_\phi
\EEA
Saddle point equations.
Deriving $G/n$ with respect to the parameters we obtain the twelve self-consistency equations
determining the order parameter values at given external pumping intensity and amount of disorder.  
\begin{itemize}
\item
Deriving w.r.t. to $Q$'s parameter we obtain the specification of Eqs. 
(\ref{f:SPeq_q}-\ref{f:tildema}) for each 1RSB replica matrix sector and for $\tilde r$ and $\tilde m$,
cf. Eqs.  (\ref{f:la_q_1rsb},\ref{f:mu_r_1rsb}),
\BEA
\lambda_{0,1}& =& \frac{\beta^2\sigma_J^2}{4} q_{0,1}|q_{0,1}|^2\ \ \ 
\\
\mu_{0,1}&=&\frac{\beta^2\sigma_J^2}{4}\left|r_{0,1}\right|^2 r_{0,1}\ \ \ 
\\
\tilde \mu&=&\frac{\beta^2\sigma_J^2}{8}\left|\tilde r\right|^2 \tilde r
\\
\nu&=&\frac{\beta J_0}{2}\left|\tilde m\right|^2 \tilde m
\EEA
\item
Deriving w.r.t  $\tilde\mu$ and $\nu$ we obtain Eqs. (\ref{f:speq_rd}),
where we define
\BEA
\langle\ldots\rangle_{\cal L}
\equiv\frac{\int_0^{2\pi}d\phi \ldots ~e^{{\cal L}(\phi;\mathbf{0},\mathbf{1})}}
{\int_0^{2\pi}d\phi~e^{{\cal L}(\phi;\mathbf{0},\mathbf{1})}}
\\
c_{\cal L}\equiv \langle\cos\phi\rangle_{\cal L}\qquad ;\qquad s_{\cal L}\equiv \langle\sin\phi\rangle_{\cal L}
\EEA
\item
Deriving $G$ w.r.t. $\lambda_{0,1}$ and $\mu_{0,1}$ and equating to zero
we obtain Eqs. (\ref{f:speq_q1_R}-\ref{f:speq_r0R}), after having integrated by part in the Gaussian measures.
To help the non-expert reader to easily derive the self-consistency equations we exemplify the calculation of Eq. (\ref{f:speq_q1_R}).
\BEA
&& 2\frac{\p G}{\p \lambda_1^R} = 0 = 1-(1-m)q_1^R
\\
\nn
&&\ \ \ -
\int {\cal D}[\mathbf{0}]\left<
\zeta_1^R c_{\cal L}+\zeta_1^I s_{\cal L}
\right>_m
/\sqrt{\Delta\lambda^R-|\Delta\mu|}
\label{f:dGdla1}
\EEA
The latter term can be simplified by integrating by part
\BEQ
\int_{-\infty}^\infty {\cal D}p(y) y  F(y) = 
\int _{-\infty}^\infty {\cal D}p(y)\frac{\p F(y)}{ \p y}
\EEQ
with $y=\zeta_1^R,\zeta_1^I$ in Eq. (\ref{f:dGdla1}), yielding
\BEA
&&\left<
\zeta_1^R c_{\cal L}+\zeta_1^I s_{\cal L}
\right>_m=\sqrt{\Delta\lambda^R-|\Delta\mu|}\times
\\
\nn
&&\quad\left<\cos^2\phi-(1-m)c_{\cal L}^2
 +\sin^2\phi-(1-m)s_{\cal L}^2
\right>_m
\EEA
The self-consistency equation can thus be rewritten as, cf. Eq. (\ref{f:speq_q1_R}),
\BEA
&&1-(1-m)q_1^R=1-(1-m)\int{\cal D}[\mathbf{0}]\left<c_{\cal L}^2+s_{\cal L}^2\right>_m
\nn
\\
&& q_1^R=\left<\left<c_{\cal L}^2\right>_m\right>_0
+\left<\left<s_{\cal L}^2\right>_m\right>_0
\EEA
Eqs. (\ref{f:speq_q0_R}-\ref{f:speq_r0R})) are analogously derived.
We notice that since from the equations $\p G/\p\lambda_{0,1}^I=0$ one obtains $q_{0,1}^I=0$
the values of the $q$ overlap are real-valued and so are the values of $\lambda$.
\end{itemize}

%\bibliography{/home/claudio/incorso/bibtex/MEGAbib}
\bibliography{MEGAbib}
\end{document}